\documentclass[a4paper,11pt]{article}
\pdfoutput=1
\usepackage{jheppub} 
\usepackage[T1]{fontenc}
\usepackage{xcolor}

\title{\boldmath Generalized holographic complexity of rotating black holes}

 \author[a,b]{Ming Zhang,}
 \affiliation[a]{Department of Physics, Jiangxi Normal University,\\ Nanchang 330022, China}
\emailAdd{mingzhang@jxnu.edu.cn}
\author[a]{Jialong Sun,}
\emailAdd{jialong-sun@jxnu.edu.cn}
\author[b,c]{Robert B. Mann}
 \affiliation[b]{Department of Physics and Astronomy, University of Waterloo,\\
Waterloo, Ontario N2L 3G1, Canada}
\affiliation[c]{Perimeter Institute for Theoretical Physics, \\ 31 Caroline St. N., Waterloo, Ontario N2L 2Y5, Canada}
\emailAdd{rbmann@uwaterloo.ca}

\abstract
{
We explore the generalized holographic complexity of odd-dimensional Myers-Perry asymptotically Anti-de Sitter (MP-AdS) black holes with equal angular momenta within the ``complexity equals anything'' proposal. We begin by determining the codimension-one generalized volume complexity by finding the extremum of the generally covariant volume functional. Locally, we show that its late-time growth rate aligns with the critical momenta associated with the extremal hypersurfaces. Globally, we discover diverse phase transitions for the complexity at early times, including first-order, second-order, and multi-critical transitions. An area law and a phase diagram are proposed to adapt to these phase behaviours, highlighting the effects of the black hole's angular momentum. At zero time, we define the generalized holographic complexity of formation and examine its scaling relations for both large near-extremal MP-AdS black holes and static charged black holes. We find that the scaling behaviours of the generalized volume complexity of formation maintain uniformity with those of the original holographic complexity formulations, except in cases where the scalar functional defining the generalized holographic complexity is infinite in the vacuum limit or at spatial infinity. Additionally, we show that these findings can be applied to codimension-zero observables.
}

\begin{document} 
\maketitle
\flushbottom

\section{Introduction}
\label{sec:intro}

Connections between quantum information and spacetime geometry continue to be an interesting avenue of research \cite{Faulkner:2022mlp,Harlow:2022qsq,deBoer:2022zka,Chen:2021lnq}. As postulated by Wheeler, the idea of ``it from bit'' \cite{Wheeler:1989ftm} can be generalized to ``it from qubit'', with  quantum information theory  playing a significant role in fundamental physics problems such as quantum gravity and quantum field theory. Indeed, entanglement \cite{Ryu:2006ef,Ryu:2006bv,Emparan:2006ni,Nishioka:2009un,Solodukhin:2011gn,Dong:2013qoa,Engelhardt:2014gca,Balasubramanian:2014sra,Almheiri:2019hni,Geng:2020qvw,Almheiri:2020cfm,Geng:2021mic} is not enough \cite{Susskind:2014moa}, as the strongly coupled dual conformal field theory (CFT) that exists at the boundary of the Anti-de Sitter (AdS) black hole thermalizes after a few AdS times \cite{Balasubramanian:2010ce,Hartman:2013qma}. 
 
Consequently, complexity has become a topic of broad interest. From the perspective of quantum information theory, there are two basic considerations \cite{Roberts:2014ifa,Aaronson:2016vto,Roberts:2016hpo,Brown:2017jil,Jefferson:2017sdb,Chapman:2017rqy,Hackl:2018ptj,Camargo:2018eof,Guo:2018kzl,Chapman:2018hou,Sinamuli:2019utz,Brandao:2019sgy,Bernamonti:2020bcf,Bhattacharyya:2021cwf,Chapman:2021jbh,Qu:2022zwq,Bhattacharyya:2023sjr,Caputa:2024vrn}. One is the complexity of a quantum state, defined by the number of (elementary) operations needed to prepare it. The other is the complexity of a unitary transformation: how many (elementary) operations are needed to apply a given unitary transformation? From a holographic perspective, the Einstein-Rosen (ER) bridge (or wormhole) in an AdS black hole has a growth rate that has prompted conjectures of its duality to the growth of complexity of the dual quantum boundary state.
 
 To measure the continuing growth of the wormhole connecting the two-sided eternal AdS black hole, holographic complexity, which is the holographic dual of circuit complexity in quantum computation theory, has emerged as a candidate quantity of interest \cite{Susskind:2014rva}. Holographic complexity has been proposed to be related to bulk spacetime geometry observables. These observables include the regularized volume of the largest codimension-one surface crossing the Einstein-Rosen bridges and extending into the black hole interior (CV proposal) \cite{Stanford:2014jda}, the low-energy action of the Wheeler-DeWitt (WDW) patch (CA proposal) \cite{Brown:2015bva, Brown:2015lvg}, as well as the volume of the WDW patch (CV 2.0 proposal) \cite{Couch:2016exn}. See \cite{Alishahiha:2015rta,Lehner:2016vdi,Carmi:2016wjl,Reynolds:2016rvl,Carmi:2017jqz,Yang:2017czx,Swingle:2017zcd,Agon:2018zso,An:2018dbz,Jiang:2018sqj,Belin:2018bpg,Goto:2018iay} for influential investigations  and \cite{Jiang:2019qea,Frassino:2019fgr,Ling:2019ien,Liu:2019mxz,Cai:2020wpc,Hernandez:2020nem,Auzzi:2022bfd,Omidi:2022whq,Erdmenger:2022lov,Zhou:2023nza,Anegawa:2023wrk,Bhattacharya:2023drv} for recent discussions on these topics. \footnote{Recently, these proposals of holographic complexity have been studied in \cite{Reynolds:2017lwq, Susskind:2021esx, Chapman:2021eyy,Jorstad:2022mls, Baiguera:2023tpt} for the (asymptotically) de Sitter (dS) spacetime according to the dS/CFT correspondence \cite{Strominger:2001pn, Strominger:2001gp, Witten:2001kn, Maldacena:2002vr}. See studies on the ``complexity = anything'' proposal for the asymptotically dS spacetime in \cite{Aguilar-Gutierrez:2023zqm, Aguilar-Gutierrez:2023pnn,Aguilar-Gutierrez:2024rka}.}

 However, these proposals all have ambiguities, either from an additional length scale (needed to obtain a dimensionless quantity from the volume) in the CV and CV 2.0 proposals, or from ambiguities in the boundary terms on null slices in the CA proposal. This, in turn, has led to the question: ``does complexity equal anything?'' \cite{Belin:2021bga}. The answer appears to be yes, insofar as there is an infinite class of diffeomorphism-invariant gravitational observables that display universal features of complexity, thus providing an infinite class of equally viable generalized holographic complexity proposals for a gravitational dual of complexity.
 
Nevertheless, generalized holographic complexity must be able to recover two key features. First, holographic complexity grows linearly in time at late times of the evolution of the entangled thermofield double state (TFD) of the CFT. Second, the growth of complexity is delayed by effects of the far-past shock wave geometry \cite{Susskind:2014jwa,Roberts:2014isa} dual to the perturbations of the TFD. It has been proposed that there can be an infinite class of covariant codimension-one \cite{Belin:2021bga} or codimension-zero \cite{Belin:2022xmt} gravitational observables of the bulk AdS black hole spacetime that exhibit these two characteristics, which are consistent with the ambiguity properties of holographic complexity. These viable candidate observables for ``complexity = anything'' have been further used to study the interior geometry of the Bañados-Teitelboim-Zanelli (BTZ) black hole \cite{Belin:2022xmt, Jiang:2023jti}, the Schwarzschild-AdS \cite{Jorstad:2023kmq} and Reissner-Nordström AdS (RN-AdS) \cite{Jorstad:2023kmq, Wang:2023eep} black holes, as well as hyperscaling-violating black branes \cite{Omidi:2022whq}, and Gauss-Bonnet geometries \cite{Wang:2023noo}. It was shown that there exist phase transitions for the generalized complexity at early times due to the introduction of a scalar function in the definition of the generalized complexity.

These studies share the common feature that the spacetimes are spherically symmetric. For the original CV and CA proposals, this led to the expectation that the complexity of formation scales with black hole entropy. However, recent studies of rotating black holes have overturned this expectation, uncovering a direct connection between the complexity of formation and thermodynamic volume for large black holes \cite{AlBalushi:2020rqe, AlBalushi:2020heq}. This finding has been replicated in other studies \cite{Andrews:2019hvq, Bernamonti:2021jyu} (see also \cite{Zhang:2022quy}). There are also developments about the scaling behaviours of the complexity for the quantum BTZ black holes \cite{Emparan:1999wa, Emparan:1999fd, Emparan:2020znc,Panella:2024sor}\footnote{See also \cite{Frassino:2022zaz,Johnson:2023dtf,Frassino:2023wpc,HosseiniMansoori:2024bfi,Feng:2024uia,Climent:2024nuj,Frassino:2024bjg}.} in \cite{Emparan:2021hyr, Chen:2023tpi}.

The ``complexity equals (almost) anything'' proposal \cite{Myers:2024vve} was raised and checked so far only for spherically symmetric black holes and three-dimensional (non-rotating or rotating) BTZ black holes. First, it is reasonable to ask whether this proposal applies to two-sided eternal higher-dimensional rotating black holes that are dual to the rotating TFDs on the spacetime boundary CFTs. Specifically, we want to check whether the late-time linear growth of the generalized holographic complexity dual to the rotating boundary states is still maintained. Second, one may wonder what new effects the rotation of the black hole will have on the early-time phase transitions of the generalized holographic complexity. Third, based on the generalized holographic complexity formulation, the generalized volume complexity of formation can be defined, and we want to check whether its scaling law  will be maintained, not only for rotating black holes but also for spherically symmetric black holes. For these reasons, it is of interest to study the ``complexity equals anything'' proposal for higher-dimensional rotating black holes.

Based on these motivations, in this paper we will study the generalized holographic complexity of odd-dimensional Myers-Perry AdS (MP-AdS) black holes \cite{Myers:1986un,Gibbons:2004js,Gibbons:2004uw} with equal angular momenta. We verify that the generalized volume complexity proposal works well for higher-dimensional MP-AdS black holes by demonstrating the late-time linear growth of the complexity. At early times, we show that there are abundant phase transition behaviours for the generalized complexity due to the nontrivial effects of black hole rotation. We will also propose a  Maxwell-like  area law for the phase transition of the complexity, which is useful for determining the phase transition points and introducing the phase diagram. At zero time, we will show that the scaling law of the generalized holographic complexity aligns with that of the original CV complexity for large near-extremal MP-AdS black holes. However, we will also show that the scaling law for the RN-AdS black hole can be affected by the specific curvature invariants chosen for the definition of the generalized holographic complexity.  Furthermore, we will explain that these results for the generalized codimension-one observables also apply to the generalized codimension-zero observables.

In section \ref{sec:mpbh}, we will provide a brief review of the odd-dimensional MP-AdS black holes with equal angular momenta. In section \ref{sec:vcf}, we will introduce the generalized volume complexity of the black hole and demonstrate that the growth rate of this complexity at late times is linear in time. In section \ref{sec:gvcse}, considering the specific form of the curvature invariant as the Gauss-Bonnet term, we will examine the specific properties of the generalized volume complexity. We will show that the generalized volume complexity undergoes several types of typical phase transitions at early times. In section \ref{jf392j293}, we will propose an  area law and a phase diagram adapting the phase transition of the generalized complexity. In section \ref{fej3i9ejfpq}, we will define the generalized complexity of formation and study its scaling laws for large MP-AdS black holes as well as spherically symmetric AdS black holes. In section \ref{jfoi39}, we will study the codimension-zero observables for the MP-AdS black holes. Section \ref{sec:clr} will be dedicated to our conclusions and discussions.

\section{MP-AdS black holes with equal angular momenta}\label{sec:mpbh}

We will study asymptotically MP-AdS black holes in $d=2N+3$ dimensions, where $N\in \mathbb{Z}^+$. This  black hole has $N+1$ independent angular momenta. To calculate  its generalized complexity, we shall take all angular momenta to be equal, thereby avoiding difficult technical issues connected with forming null hypersurfaces in such spacetimes \cite{AlBalushi:2019obu,Imseis:2020vsw}.
This leads to the reduction of the original cohomogeneity-$(N+1)$ MP-AdS metric with an isometry group of $\mathbb{R} \times \mathrm{U}(1)^{N+1}$ to one with an enhanced isometry of $\mathbb{R} \times \mathrm{U}(1) \times \mathrm{SU}(N+1)$, as given by \cite{Myers:1986un,Gibbons:2004js,Gibbons:2004uw,Kunduri:2006qa}
\begin{equation}\label{mpadsmet}
\mathrm{d}s^2 = -f(r)^2 \mathrm{d}t^2 + g(r)^2 \mathrm{d}r^2 + h(r)^2[\mathrm{d}\psi + A_{(N)} - \bar{\Omega}(r) \mathrm{d}t]^2 + r^2 \hat{g}_{ab} \mathrm{d}x^a \mathrm{d}x^b\,,
\end{equation}
where
\begin{align}
f(r) &= \frac{r}{g(r) h(r)}\,, \\
g(r) &= \left(1 + \frac{r^2}{\ell^2} - \frac{2m\Xi}{r^{2N}} + \frac{2ma^2}{r^{2N+2}}\right)^{-1/2}\,, \\
h(r) &= r \left(1 + \frac{2ma^2}{r^{2N+2}}\right)^{1/2}\,, \\
\bar{\Omega}(r) &= \frac{2ma}{r^{2N} h^2}, \quad \Xi = 1 - \frac{a^2}{\ell^2}\,.
\end{align}
Here, $m$, $a$, and $\ell$ have their usual meanings. In \eqref{mpadsmet}, $\hat{g}_{ab}$ is the Fubini-Study construction of the Einstein-Kähler metric on $\mathbb{CP}^N$, and $A_{(N)}$ is the $\mathbb{CP}^N$ Kähler potential, given by \cite{Hoxha:2000jf,Dias:2010eu}
\begin{align}
\hat{g}_{ab} \mathrm{d}x^a \mathrm{d}x^b &\equiv \mathrm{d}\Sigma_N^2 = \frac{\mathrm{d}R_N^2}{(1+R_N^2)^2} + \frac{1}{4}\frac{R_N^2}{(1+R_N^2)^2}(\mathrm{d}\Psi_N + 2A_{(N-1)})^2 + \frac{R_N^2}{1+R_N^2}\mathrm{d}\Sigma_{N-1}^2\,, \\
A_{(N)} &= \frac{1}{2}\frac{R_N^2}{1+R_N^2}(\mathrm{d}\Psi_N + 2A_{(N-1)})\,,
\end{align}
where the coordinates satisfy $R_N \geq 0$ and $0 \leq \Psi_N \leq 4\pi$. Specifically, $\mathbb{CP}^1$ is isomorphic to $S^2$, in which case we have
\begin{align}
\mathrm{d}\Sigma_1^2 &= \frac{1}{4}(\mathrm{d}\theta^2 + \sin^2\theta \mathrm{d}\phi^2)\,, \\
A_{(1)} &= \frac{1}{2}\cos\theta \mathrm{d}\phi\,.
\end{align}
One can verify that $\operatorname{Ric}(\hat{g}) = 2(N+1)\hat{g}$. The conserved Kähler form on $\mathbb{CP}^N$, defined by $J_N = \mathrm{d}A_N/2$, satisfies the relation $J_a^b J_{bc} = -\hat{g}_{ac}$.

The null hypersurface of the odd-dimensional MP-AdS black hole with equal angular momenta  can be obtained in a similar manner to the static cases \cite{AlBalushi:2020rqe,AlBalushi:2020heq}, unlike the general rotating ones \cite{AlBalushi:2019obu,Imseis:2020vsw}. Specifically, we can define the infalling Eddington-Finkelstein coordinate as
\begin{equation}\label{iefc}
v = t + r^*(r, \psi_i)\,,
\end{equation}
where $r^*$ is the tortoise coordinate, and $\psi_i$ are the angular coordinates. Following the procedures conducted in \cite{AlBalushi:2020rqe,AlBalushi:2019obu}, we find that
\begin{equation}
\frac{\mathrm{d}r^*}{\mathrm{d}r} = \frac{g(r)}{f(r)} = \frac{g(r)^2 h(r)}{r}\,,
\end{equation}
which implies that $r^*$ is independent of $\psi_i$. Furthermore, we will need to determine the event horizon $r_+$ and the Cauchy horizon $r_-$ of the MP-AdS black hole, which can be obtained by solving $g(r)^{-2} = 0$.

By setting $m=0$, we can get the AdS vacuum; for $d=5$, the explicit form of the metric reads
\begin{equation}
   \mathrm{d}s^2_{\mathrm{AdS}}= -\left(1+\frac{r^2}{\ell^2}\right) \mathrm{d} t^2+\left(1+\frac{r^2}{\ell^2}\right)^{-1} \mathrm{d} r^2+\frac{r^2}{4} \mathrm{d} \theta^2+\frac{r^2}{4} \mathrm{d} \phi^2+r^2 \mathrm{d} \psi^2+r^2\cos\theta \mathrm{d} \phi \mathrm{d} \psi\,.
\end{equation}
Its determinant is $-r^6\sin^6\theta/16$, which is the same with that of the metric \eqref{mpadsmet}.

\begin{figure}[t!]
	\centering
	\includegraphics[width=2.7in]{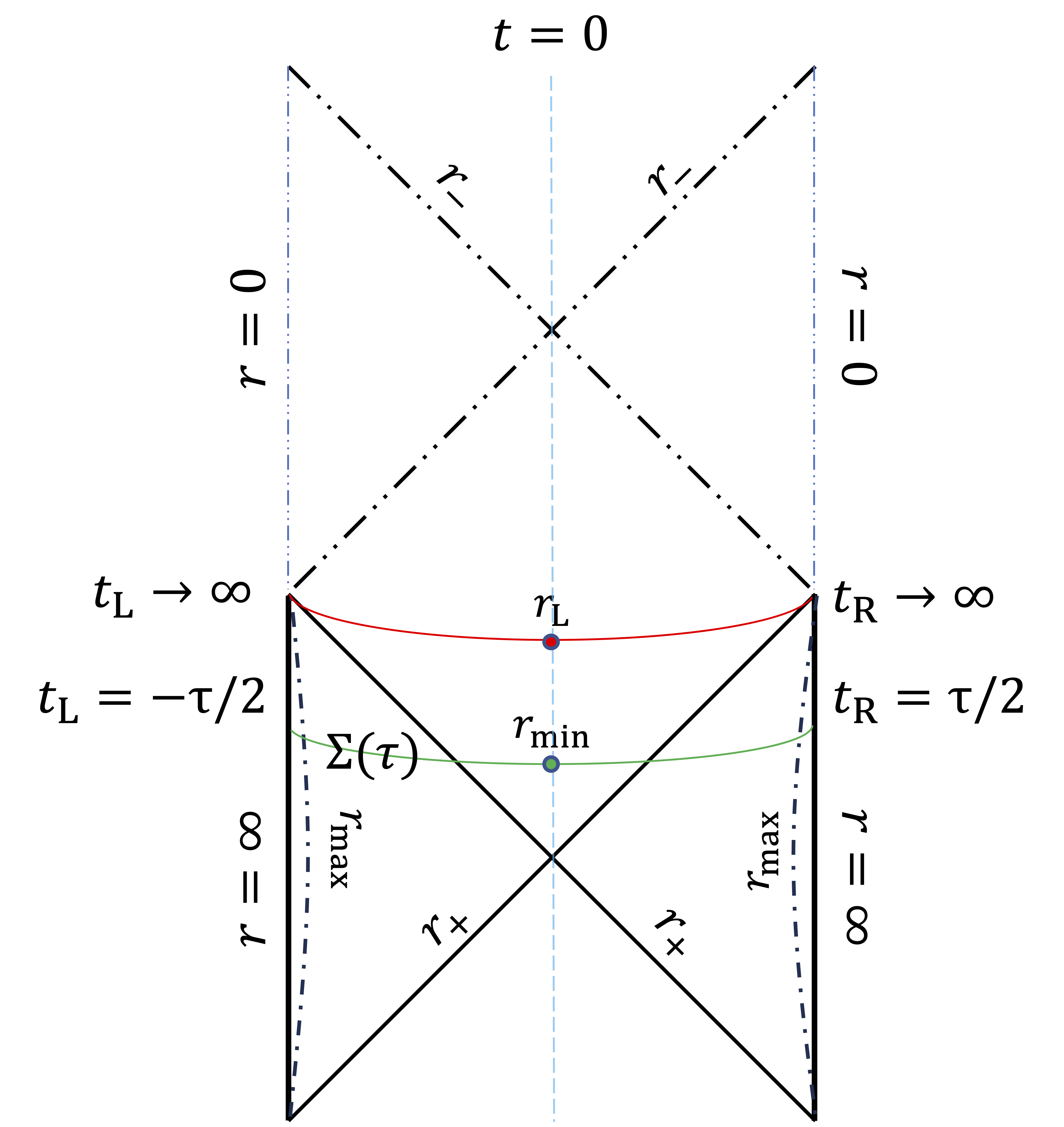}
	\caption{The Penrose diagram of the MP-AdS black hole. $\Sigma(\tau)$ represents an extremal hypersurface 
 (shown in green) at the boundary time $t_{\mathrm{R}}=-t_{\mathrm{L}}=t$. $r_{\mathrm{L}}$ is a constant-$r$ hypersurface (shown in red) that can be obtained using \eqref{conmu}. $r_{\min}$ is the turning point that satisfies \eqref{urormin}. The vertical blue line in the middle represents $t=0$. $r_{\max}$ is the ultraviolet (UV) cutoff surface that is used in \eqref{bdtau}.}\label{pendg}
\end{figure}

\section{Generalized volume complexity of MP-AdS black holes}

\subsection{Growth rate at late time}\label{sec:vcf}

As advocated in \cite{Belin:2021bga}, we can define a codimension-one observable $\mathcal{V}$
\begin{equation}\label{codoneo} 
 \max_{\partial \Sigma=\Sigma_{\mathrm{CFT}}} \frac{\mathcal{V}}{\mathrm{G}_\mathrm{N} \ell}
\equiv \max_{\partial \Sigma (\tau)=\Sigma_{\mathrm{CFT}}} \left[\frac{1}{\mathrm{G}_\mathrm{N} \ell} \int_{\Sigma (\tau)} \mathrm{d}^{d-1} x \sqrt{\tilde{h}} F(g_{\mu \nu}, \mathcal{R}_{\mu \nu \rho \sigma}, \nabla_\mu)\right] 
=C(\Sigma_{\mathrm{CFT}}) \,,
\end{equation}
where the latter relation
generalizes the original CV conjecture,
with $\tilde{h}_{ab}$ the induced metric on the bulk hypersurface $\Sigma (\tau)$ (see Fig. \ref{pendg}) and $\tilde{h}$ its determinant. It means that the generalized complexity of the TFD boundary states is measured by the covariant gravitational observable 
$F(g_{\mu \nu}, \mathcal{R}_{\mu \nu \rho \sigma}, \nabla_\mu)>0$
extremized over all spacelike bulk hypersurfaces $\Sigma(\tau)$ that are anchored to the boundary by the condition $\partial \Sigma (\tau)=\Sigma_{\mathrm{CFT}}$, where $\Sigma_{\mathrm{CFT}}$ is a constant time slice in the boundary CFT. As suggested by \cite{Belin:2021bga}, we can define the scalar function $F$ using the curvature invariants of the bulk MP-AdS geometry, such as $\mathcal{R}_{a b c d} \mathcal{R}^{a b c d}$, $\mathcal{R}_{a b} \mathcal{R}^{a b}$, and $C_{abcd}C^{abcd}$ \cite{Kraniotis:2021qah}. 
Here we choose $F$ itself to determine the extremal surface while also evaluating the same function $F$ on that extremal surface for simplicity
  \cite{Belin:2021bga}.   The two decoupled CFTs on the boundary, which are dual to the two-sided bulk geometry, are rotating entangled TFD states given by
\begin{equation}
|\mathrm{rTFD}\rangle = \frac{1}{\sqrt{Z(\beta, \Omega)}} \sum_n e^{-\beta(E_n+\Omega J_n)/2} e^{-i(E_n+\Omega J_n)(t_{\mathrm{L}}+t_{{\mathrm{R}}})/2} |E_n, J_n\rangle_{{\mathrm{L}}} |E_n, J_n\rangle_{{\mathrm{R}}}\,,
\end{equation}
where $\beta, E_n, J_n, \Omega$ are the inverse of the temperature, the energy, angular momentum, angular velocity of the states $|E_n, J_n\rangle_L$, $|E_n, J_n\rangle_R$ on the left and right copies of CFTs, respectively. $Z(\beta,\Omega)=\sum_n e^{-\beta(E_n+\Omega J_n)/2}$ is the partition function \cite{Maldacena:2001kr,Hartman:2013qma,Israel:1976ur,Bernamonti:2021jyu}. To evolve the Hamiltonian, we can choose symmetric boundary time slices $t_{{\mathrm{R}}}=-t_{{\mathrm{L}}}=t$, as shown in Fig. \ref{pendg}.

Let us proceed without specifying the scalar function $F$. By evaluating the curvature invariants (see Appendix \ref{app:cio}), we know that $F$ is independent of the angular coordinates, i.e., $F=F(r)$. Consequently, it is useful to rewrite the metric \eqref{mpadsmet}  in  half-null coordinates \eqref{iefc}, with
$x^\mu=(v, r, \vec{\Omega})$ and the spacelike hypersurfaces $\Sigma(\tau)$   parametrized by $(v(\sigma), r(\sigma), \vec{\Omega})$. Here, $\vec{\Omega}$ represents the angular coordinates, and $\sigma$ can be viewed as a radial coordinate of the hypersurface $\Sigma(\tau)$. As a result,  the generalized volume functional in \eqref{codoneo} becomes
\begin{equation}\label{volfunc}
\begin{aligned}
\mathcal{V} &= \int_{\Sigma} \mathrm{d}^{d-1} x \sqrt{\tilde{h}} F\left(g_{\mu \nu}, \mathcal{R}_{\mu \nu \rho \sigma}, \nabla_\mu\right) \\
&= {\Omega}_{d-2} \int_{\Sigma} h(r) r^{d-3} \sqrt{-f(r)^2 \dot{v}^2+2 g(r) f(r) \dot{v} \dot{r}} \mathfrak{a}(r) \mathrm{d} \sigma\,,
\end{aligned}
\end{equation}
where ${\Omega}_{d-2}=2\pi^{(d-1)/2}/\Gamma((d-1)/2)$ is the volume of the $(d-2)$-dimensional sphere and the scalar factor $\mathfrak{a}(r)$ is obtained by evaluating $F$ using the odd-dimensional bulk MP-AdS geometry with equal angular momenta. The overdot denotes the derivative with respect to the coordinate $\sigma$.

Thus, finding the extremal value of the volume functional \eqref{volfunc} over all the spacelike bulk hypersurfaces $\Sigma(\tau)$ is equivalent to solving a classical mechanics problem with the Lagrangian
\begin{equation}\label{fj3ei9p28p}
\mathcal{L} = h(r) r^{d-3} \sqrt{-f(r)^2 \dot{v}^2+2 g(r) f(r) \dot{v} \dot{r}} \mathfrak{a}(r)\,.
\end{equation}
It is evident that there is a conserved momentum $P_v(\tau)$ conjugate to the ingoing coordinate $v$, which reads
\begin{equation}\label{constc2}
P_v=\frac{\partial \mathcal{L}_{\mathrm{}}}{\partial \dot{v}}=\frac{\mathfrak{a}(r) h(r) r^{d-3}\left( g(r)f(r)\dot{r}-f(r)^2 \dot{v}  \right)}{\sqrt{-f(r)^2 \dot{v}^2+2 g(r) f(r) \dot{v} \dot{r}} }=g(r)f(r)\dot{r}-f(r)^2 \dot{v}\,,
\end{equation}
where in the last step we have chosen a gauge-fixing condition
\begin{equation}\label{constc1}
\sqrt{-f(r)^2 \dot{v}^2+2 g(r) f(r) \dot{v} \dot{r}} = h(r) r^{d-3} \mathfrak{a}(r)
\end{equation}
due to the diffeomorphism invariant nature of the observable \eqref{codoneo}. Thus, by applying \eqref{constc2} together with \eqref{constc1}, we have
\begin{align}
\dot{r} &= \pm \sqrt{\frac{r^{2 d-6} h(r)^2 \mathfrak{a}(r)^2}{g(r)^2}+\frac{P_v^2}{f(r)^2 g(r)^2}} \label{dotr8}\,, \\
\dot{v} &= \frac{1}{f(r)^2}\left(-P_v \pm \sqrt{\mathfrak{a}(r)^2 r^{2 d-6} f(r)^2 h(r)^2+P_v^2}\right)\,. \label{dotr82}
\end{align}
By eliminating the parameter $\sigma$ in \eqref{dotr8} and \eqref{dotr82}, we can obtain the profile of the hypersurface $r(v)$ that extremizes the volume functional \eqref{volfunc} as well as the codimension-one observable \eqref{codoneo}. 

Furthermore, the radial profile equation \eqref{dotr8} can be recast as
\begin{equation}\label{wtepid}
f(r)^2 g(r)^2 \dot{r}^2+U_0(r)=P_v^2\,,
\end{equation}
where we have defined 
\begin{equation}\label{pida}
U_0 (r)=-r^{2 d-6} h(r)^2 \mathfrak{a}(r)^2 f(r)^2\,,
\end{equation}
which can be viewed as the effective potential of the classical mechanics problem with the Lagrangian \eqref{fj3ei9p28p}.

For the non-relativistic classical mechanics problem \eqref{wtepid}, it is evident that the effective potential vanishes at both the inner and outer horizons $r_{\mp}$. As a result, there will be at least one extremal point between the two horizons
where $\dot{r}=0$. For the MP-AdS black hole, as we will show later, there can be more than one local extremal points for the effective potential \eqref{pida}. The local maximum extremal point $r=r_f$ can be obtained by solving
\begin{equation}\label{conmu}
U_0\left(r_f\right)=P_{\infty}^2\,, \quad U_0^{\prime}\left(r_f\right)=0\,, \quad U_0^{\prime \prime}\left(r_f\right) \leq 0\,,
\end{equation}
where we have defined the critical conserved momentum at $r_f$ as $P_\infty$. 

From \eqref{constc2} and \eqref{constc1} we can obtain
\begin{equation}\label{dvdr1}
\frac{\mathrm{d}v}{\mathrm{d}r}=\mp \frac{g(r)P_v}{f(r) \sqrt{P_v^2-U_0(r)}} \pm \frac{g(r)}{f(r)}\,.
\end{equation}
From \eqref{iefc} we have 
\begin{equation}\label{dvdr2}
\frac{\mathrm{d}v}{\mathrm{d}r}=\frac{\mathrm{d}t}{\mathrm{d}r}+\frac{g(r)}{f(r)}
\end{equation}
for the ingoing coordinate $v$. 
Comparing \eqref{dvdr1} with \eqref{dvdr2}, we have
\begin{equation}\label{jfi9292}
\frac{\mathrm{d}t}{\mathrm{d}r}=- \frac{g(r)P_v}{f(r) \sqrt{P_v^2-U_0(r)}}\,,
\end{equation}
yielding the boundary time
\begin{equation}\label{bdtau}
\tau \equiv 2 t_{\mathrm{R}}=-2 \int_{r_{\min }}^{r_{\max}} \mathrm{d} r \frac{g(r) P_v}{f(r) \sqrt{P_v^2-U_0(r)}}\,,
\end{equation}
where $r_{\max}$ is the UV cutoff surface and $r_{\min}$ is the turning point of the pseudo non-relativistic particle determined by
\begin{equation}\label{urormin}
U_0 \left(r_{\min }\right)=P_v^2\,,
\end{equation}
which is reduced from \eqref{wtepid}. When $P_v=P_\infty$, the solution of \eqref{urormin} as a turning point may or may not be an extremal point: if the second and third conditions in \eqref{conmu} are satisfied, we have an extremal point $r_{\min }=r_f$; otherwise, the turning point is not an extremal point, i.e., $r_{\min }\neq r_f$.

Moreover, by substituting \eqref{constc1} and \eqref{dotr8} into \eqref{volfunc}, we can get the extremal value of the generalized volume functional as
\begin{equation}\label{vmax2}
\mathcal{V}_{\max}=2\Omega_{d-2} \int_{r_{\min }}^{r_{\max}} \mathrm{d} r \frac{\mathfrak{a}(r)^2h(r)^2 r^{2d-6}f(r) g(r)}{\sqrt{P_v^2-U_0(r)}}\,.
\end{equation}
Furthermore, employing \eqref{pida}, \eqref{jfi9292}, and \eqref{bdtau}, \eqref{vmax2} can be reformulated as \cite{Belin:2022xmt}
\begin{equation}\label{jfq9348}
   \mathcal{V}_{\max}= \Omega_{d-2}\int_0^{\tau}\frac{U_0\left[r(\tau^\prime)\right]}{P_v(\tau^\prime)}\mathrm{d}\tau^\prime.
\end{equation}
Here $P_v(\tau^\prime)$ denotes that with a given  conserved momentum $P_v\neq P_\infty$, we may have a corresponding boundary time $\tau<\infty$ given by \eqref{bdtau}; if $P_v\to P_\infty$, we have $\tau\to\infty$, which cooresponds to either an extremal surface $r\to r_f$ satisfying \eqref{conmu} or a turning point $r_{\min}$ satisfying \eqref{urormin}. (cf. Fig. 8 in \cite{Belin:2022xmt}.)

For the generalized volume complexity defined in \eqref{codoneo}, from \eqref{jfq9348} it is not difficult to show that (see also Appendix \ref{dcdtaud})
\begin{align}\label{djio289}
\frac{\mathrm{d} C}{\mathrm{d} \tau} =\frac{\Omega_{d-2}}{\mathrm{G}_\mathrm{N} \ell}P_v (\tau)\,.
\end{align}
At late times, this implies that the generalized codimension-one CV complexity increases linearly with the boundary time as
\begin{equation}\label{vjiq398}
\lim _{\tau \rightarrow \infty}\left(\frac{\mathrm{d} C}{\mathrm{d} \tau}\right)=\frac{\Omega_{d-2}}{\mathrm{G}_\mathrm{N} \ell} P_\infty\,,
\end{equation}
where the critical momentum $P_\infty$ is determined by \eqref{conmu}. We note that there may be more than one critical momenta (for which we will show later) in certain conditions. Each local critical momentum corresponds to an independent phase where the generalized CV complexity exhibits late-time linear growth.   This makes that it is necessary to determine the early time behaviour of the complexity from a global viewpoint.

\subsection{Growth rate at early time}\label{sec:gvcse}

In order to elucidate the global properties of the generalized volume complexity, we will now assign an explicit expression for the scalar function $F$ in \eqref{codoneo}. Specifically, we choose the scalar function $F$ to be
\begin{equation}\label{fonelg}
F=\mathfrak{a}(r)=1+\lambda \ell^4 \mathcal{G}\,,
\end{equation}
where $\lambda$ represents a dimensionless coupling constant, and the higher curvature term
\begin{equation}\label{fji933498}
\mathcal{G}=\mathcal{R}^2-4 \mathcal{R}_\nu^\mu \mathcal{R}_\mu^\nu+\mathcal{R}_{\rho \kappa}^{\mu \nu} \mathcal{R}_{\mu \nu}^{\rho \kappa}
\end{equation}
is the Gauss-Bonnet invariant, where $\mathcal{R}, \mathcal{R}_{\mu\nu}, \mathcal{R}_{\mu\nu\rho\kappa}$ are the curvature scalar, Ricci tensor, and Riemann tensor of the background MP-AdS black hole, respectively. $\mathcal{G}$ is a topological term for $d=4$. The specific values for the five-dimensional and seven-dimensional curvature invariants are provided in Appendix \ref{app:cio}.  This choice is equivalent to choosing the square of the Weyl tensor
\cite{Jorstad:2023kmq,Wang:2023eep} in vacuum spacetimes up to an additive constant.
We will now study the generalized volume complexity of the MP-AdS black hole, focusing primarily on the five-dimensional case, as other cases will exhibit similar characteristics. See Appendix \ref{app:seven} for an example of the seven-dimensional case.

For different sets of values of the black hole parameters $m, a, \ell$, and the coupling parameter $\lambda$, we find that there can be up to four local maxima   for the effective potential \eqref{pida} specified by the scalar function \eqref{fonelg} between the inner and outer horizons of the five-dimensional MP-AdS black hole\footnote{Due to the highly nonlinear dependence of the effective potential  on the parameters, it is inefficient to determine which sets of parameters yield specific patterns of the effective potentials.}. This is also the case for other odd-dimensional scenarios. The variations of the complexity with respect to time will be similar to those of the Schwarzschild-AdS \cite{Jorstad:2023kmq} or RN-AdS black holes \cite{Wang:2023eep} if there are one or two local maxima in the effective potential\footnote{In the higher-dimensional cases, the number of maxima  remains up to one and two for the Schwarzschild-AdS and RN-AdS black holes, respectively.}. We will focus our study on the case with three local maxima, as the variations of the complexity in this situation are notably distinct from the previous spherically symmetric cases. We have found that cases with four maxima are  similar to those with three maxima; therefore we will not discuss them further, apart from Appendix \ref{app:seven}, where we give an example of the four maxima case for the seven-dimensional MP-AdS black hole.  

By using \eqref{codoneo}, \eqref{bdtau}, and \eqref{vmax2}, we can respectively calculate the boundary time in terms of the conserved momentum and the time-variation of the generalized volume complexity. The variations of the boundary time $\tau$ with respect to the conserved momenta are depicted in the middle diagrams of Figs. \ref{effpmp1}, \ref{effpmp2}, \ref{effpmp3}, and \ref{effpmp4}. We observe that the boundary time diverges for critical momenta, which correspond to the locally maximal extremal points $r_L$, $r_M$, and $r_R$ in the left diagrams of Figs. \ref{effpmp1}, \ref{effpmp2}, \ref{effpmp3}, and \ref{effpmp4}. These critical momenta can be obtained from \eqref{conmu} as $P_L=\sqrt{U_0 (r_L)}, P_M=\sqrt{U_0 (r_M)}, P_R=\sqrt{U_0 (r_R)}$. Around each critical momentum there is one branch with a positive slope (on the left) and another with a negative slope (on the right). Branches with positive slopes can contribute as  possible real paths for the complexity variations in time, as illustrated by the colored curves in the right diagrams of Figs. \ref{effpmp1}, \ref{effpmp2}, \ref{effpmp3}, and \ref{effpmp4}. However,   branches with  negative slope do not contribute to the real path of the complexity variation, as the corresponding complexity is always less than that for the counterparts with positive slopes. Correspondingly, we can observe in the right diagrams of Figs. \ref{effpmp1}, \ref{effpmp2}, \ref{effpmp3}, and \ref{effpmp4} that the complexity variation curves also split into three unconnected branches.

In general, we have identified several types of effective potentials that result in different  configurations of the generalized volume complexity. Essentially, we can classify them into four types:
\begin{itemize}
\item Type (A): The generalized volume complexity chronologically begins from phase 1 and ends at phase 3, as illustrated in Fig. \ref{effpmp1}, where we have plotted the region between the inner and outer horizons. We observe that there are three local maxima  $r_L,\, r_M,\,r_R$, with the values of the effective potential satisfying $U_0(r_L)>U_0(r_M)>U_0(r_R)$. According to \eqref{conmu},   the critical conserved momenta at each extremal point comply with the relation $P_{L}>P_{M}>P_{R}$. We have chosen the parameters as $m=51/50,\; a=23/50, \;\ell=10,\; \lambda=1/10^{339/50}$, such that $P_{L}\approx 2.460, \;P_{M}\approx 1.890,\; P_{R}\approx 0.859$. From \eqref{bdtau}, the boundary time at each critical conserved momentum diverges, as shown in the middle diagram of Fig. \ref{effpmp1}. According to the definition \eqref{codoneo} for the generalized volume complexity, if there are several spacelike bulk hypersurfaces $\Sigma (\tau)$ that extremize the volume functional \eqref{volfunc}, the generalized holographic complexity of the TFD boundary states tends to be the one with maximal generalized volume \eqref{volfunc}. This is similar to a system thermodynamically tending to a state with lower free energy.  Despite there being several possible  
trajectories for the   generalized volume complexity
(as shown  in the right diagram of Fig. \ref{effpmp1}),
the complexity evolves in a manner that ensures its maximal growth. That is, the complexity begins from phase 1, which corresponds to the phase with the momentum $P<P_R$ and are  blue curves in the middle and right panels of  Fig. \ref{effpmp1}.  As $\tau$ increases, a point is reached at which the generalized volume is no longer maximal, and the trajectory continues along the blue dot-dashed curve in the right panel of Fig. \ref{effpmp1}, until the point $r_R$ is reached coming from the right, where $\tau$ diverges. Continuing further, $\tau$ decreases and $P_v$ increases, yielding the trajectory given by the lowest black trajectory in the right panel of Fig. \ref{effpmp1}. Once the minimal value of $\tau $ is reached (at $P_v\approx 1.825$), the leftmost cusp in the trajectory in the right panel of  Fig. \ref{effpmp1} is reached and the trajectory reverses direction, again out to infinite $\tau$ where the point $r_M$ is attained. The process then repeats, but this time after the cusp is reached the trajectory in the right panel intersects the earliest trajectory (at the blue point) and the complexity follows the red curve.  The net effect is that the complexity begins from phase 1 (the solid blue curve) and then at a certain time $\tau=\tau_{\text{transition}}\approx 2.32$, changes to phase 3 (the solid red curve), which corresponds to the phase with momentum $P_M<P\lesssim P_L$. All of phase 2, which corresponds to   momenta around $P_M$, is suppressed.

\begin{figure}[t!]
	\centering
	\includegraphics[width=\textwidth]{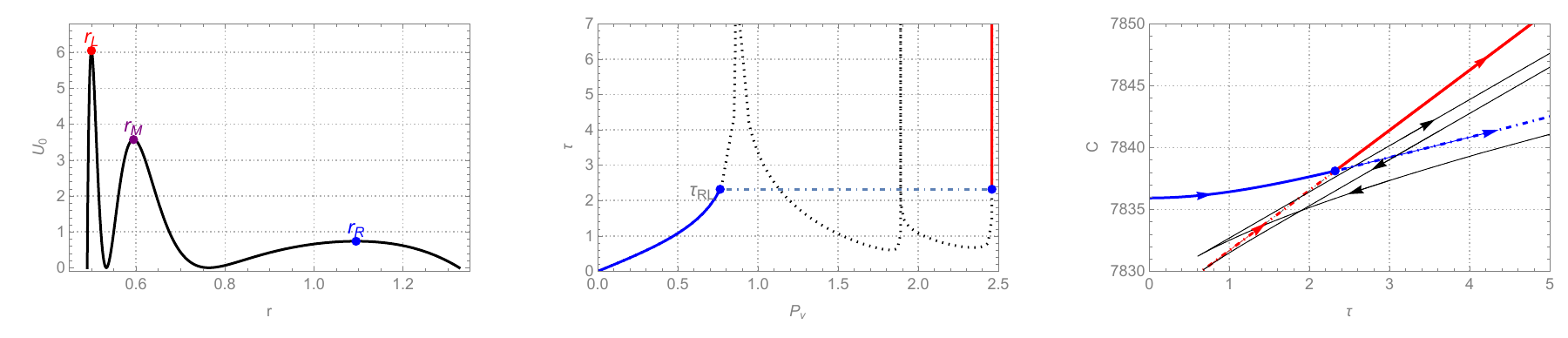}
	\caption{Left: the effective potential $U_0$ between the inner and outer horizons. The effective potential has local extrema at $r=r_L, r_M, r_R$ (this applies to all subsequent cases). Middle: the variation of the boundary time with respect to the conserved momentum. At the time $\tau=\tau_{\text{RL}}$, the boundary time undergoes a sudden change from the blue branch (phase 1) to the red one (phase 3). Right: the variation of the generalized volume complexity $C$ in \eqref{codoneo} with respect to the boundary time $\tau$, with $F$ given by \eqref{fonelg}. Arrows indicate the trajectories following the curves in the middle figure, with the black lines corresponding to the dotted black curves. At $\tau=\tau_{\text{RL}}$, the complexity experiences a sudden change in its growth rate from phase 1 to phase 3. We have set $m=51/50,\; a=23/50,\; \ell=10,\; \lambda=1/10^{339/50}$ for the five-dimensional MP-AdS black hole.}\label{effpmp1}
\end{figure}

\begin{figure}[t!]
	\centering
	\includegraphics[width=\textwidth]{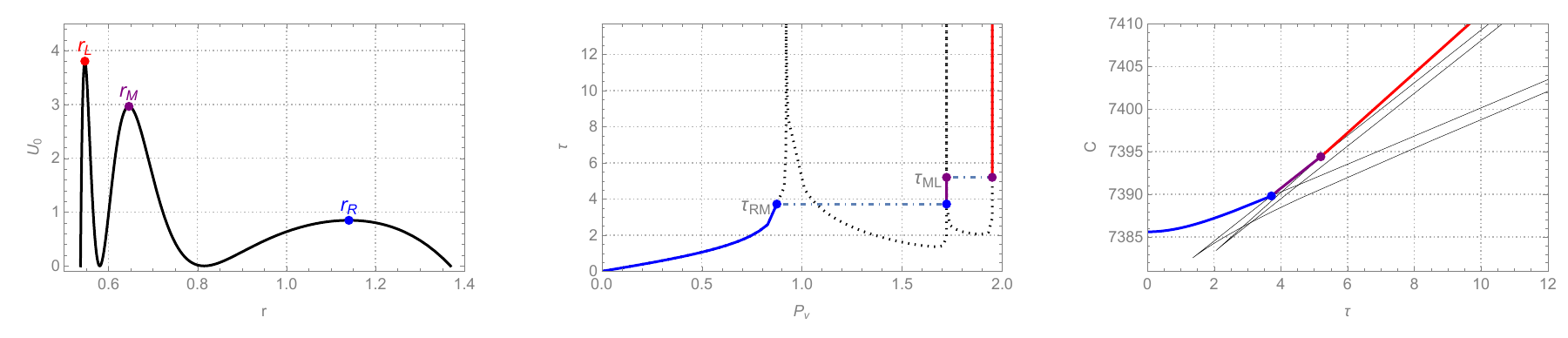}
	\caption{Left: the effective potential $U_0$ between the inner and outer horizons. Middle: the variation of the boundary time with respect to the conserved momentum. The boundary time undergoes a sudden shift from the blue branch (phase 1) to the purple branch (phase 2), and then further to the red branch (phase 3). Right: the variation of the generalized volume complexity with respect to the boundary time. At $\tau=\tau_{\text{RM}}, \tau_{\text{ML}}$, the complexities experience  sudden changes in growth rate, transiting from phase 1 to phase 2 and from phase 2 to phase 3, respectively. For this illustration, we set $m=11/10,\; a=1/2,\; \ell=11, \;\lambda=1/10^{341/50}$ for the five-dimensional MP-AdS black hole.
Black lines in the right diagram correspond to the dotted curve in the middle figure; arrows have been suppressed. 
 }\label{effpmp2}
\end{figure}

\begin{figure}[t!]
	\centering
	\includegraphics[width=\textwidth]{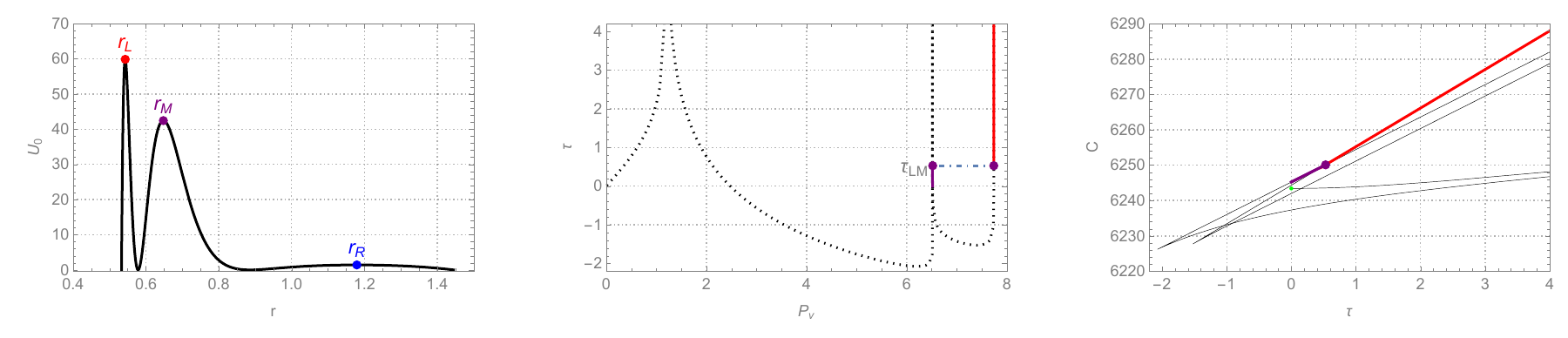}
	\caption{Left: the effective potential $U_0$ between the inner and outer horizons. Middle: the variation of the boundary time with respect to the conserved momentum. The boundary time undergoes a sudden shift from the purple branch (phase 2) to the red branch (phase 3). Right: the variation of the generalized volume complexity with respect to the boundary time. We can read the figure from the green point where $\tau=0$. The complexity experiences a sudden change in its growth rate, transiting from phase 1 to phase 2 at $\tau=\tau_{\text{LM}}$. We have set $m=6/5,\; a=1/2,\; \ell=14, \;\lambda=1/10^{341/50}$ for the five-dimensional MP-AdS black hole.  Black lines in the right diagram correspond to the dotted curves in the middle figure; arrows have been suppressed.  }\label{effpmp3}
\end{figure}

\begin{figure}[t!]
	\centering
	\includegraphics[width=\textwidth]{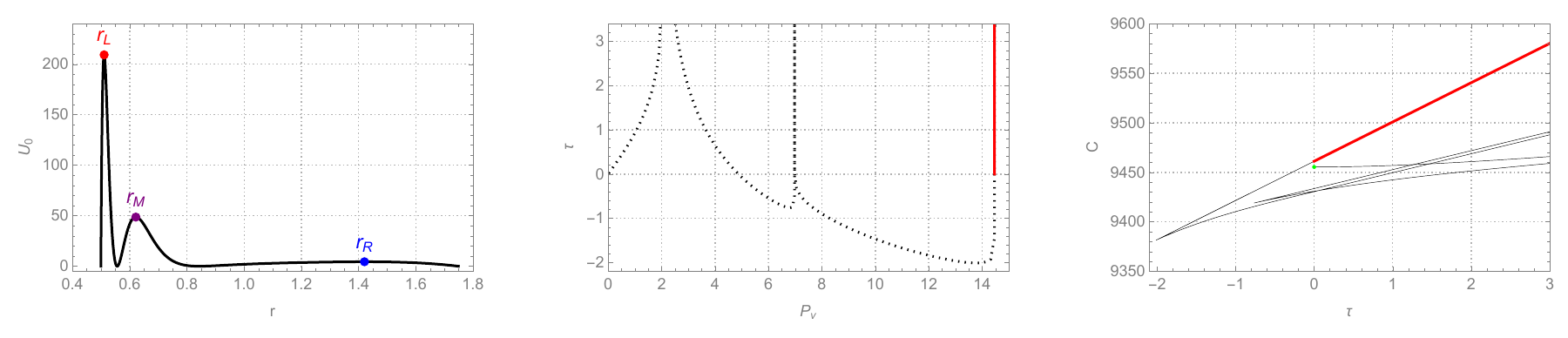}
	\caption{Left: the effective potential $U_0$ between the inner and outer horizons.  Middle: the variation of the boundary time with respect to the conserved momentum, with the real path for the time variation represented by the red curve. Right: the variation of the generalized volume complexity with respect to the boundary time, where there is only one phase 3 (represented by the red curve) for the complexity. We can read the figure from the green point where $\tau=0$. The parameters are set as $m=44/25,\; a=12/25,\; \ell=36/5, \;\lambda=1/10^{31/5}$ for the five-dimensional MP-AdS black hole.  Black lines in the right diagram correspond to the dotted curve in the middle figure; arrows have been suppressed.   }\label{effpmp4}
\end{figure}

\item Type (B): With a slight change of parameters in the type (A), we find that phase 2 (the purple curve) can appear during intermediate values of $\tau$, with the trajectory going from phase 1 (the blue curve) to phase 2 to phase 3 (the red curve).  The reasoning is the same as in the previous case, with the final curve obtained by tracing out the trajectory of maximal volume.
This is shown in Fig. \ref{effpmp2}, where $m=44/25, \;a=12/25,\; \ell=36/5,\; \lambda=1/10^{31/5}$.   The transition time from phases 1 to 2 is $\tau=\tau_{\text{transition}}\approx 3.72$ and the transition time from phases 2 to 3 is $\tau=\tau_{\text{transition}}\approx 5.72$. Qualitatively, we can find that if the differences of extremal values of the effective potentials are not significant, i.e., $U_{0}(r_L) \gtrsim U_{0}(r_M)\gtrsim U_{0}(r_R)$, then we may have this class of complexity phase transitions.

\item Type (C): In this case phase 1 (the phase with the momentum $P<P_R$) is never the curve of maximal  generalized volume, and the  complexity begins at phase 2 (the purple curve) and ends at phase 3 (the red curve). This is shown in Fig. \ref{effpmp3} where we can observe that there are three maximal   points for the effective potential between the inner and outer horizons, with values  $U_{0}(r_L)\approx 59.892,\; U_{0}(r_M)\approx 42.454,\; U_{0}(r_R)\approx 1.412$. 
The complexity grows from $\tau=0$ to $\tau=\tau_{\text{transition}}\approx 0.53$ and jumps from phase 2 to phase 3.   

\item Type (D): Only phase 3 (the red curve) is present in this case, as shown in Fig. \ref{effpmp4}. This means that there will be no phase transition for the generalized volume complexity, even if there is  more than one locally maximal extremal point for the effective potential. This is quite novel in itself and  was not reported in \cite{Wang:2023eep} for the RN-AdS black hole, though there can also be only one phase even for a configuration where the left extremal value is greater than the right extremal value of the effective potential. For the parameter we choose in Fig. \ref{effpmp4}, we have $U_{0}(r_L)\approx 209.244,\; U_{0}(r_M)\approx 48.655,\; U_{0}(r_R)\approx 4.444$. 
\end{itemize}

One point worth noting is that in Figs. \ref{effpmp3} and \ref{effpmp4}, the generalized volume complexity varies at  negative boundary times. According to the discrete symmetry $(t, \psi_i) \rightarrow (-t, -\psi_i)$, complexity at these times can be attributed to the corresponding positive boundary time. Subsequently, based on the definition \eqref{codoneo} of complexity, which refers to the maximum volume among all possible volumes of the codimension-one hypersurfaces,   complexity at the negative boundary time does not constitute an actual path of complexity variation.

A couple of additional points should be noted: if $U_0(r_L)<U_0(r_M)<U_0(r_R)$ or $U_0(r_M)<U_0(r_L)<U_0(r_R)$, then the configuration of the complexity variation resembles the Schwarzschild-AdS case, where the generalized complexity increases monotonically to infinity; on the other hand, if $U_0(r_L)>U_0(r_R)>U_0(r_M)$, the configuration of the complexity variation is akin to the RN-AdS case. Moreover, in the limit of vanishing angular momenta for the MP-AdS black hole, the configurations of the complexity variation reduce to the  Schwarzschild-AdS case.

From the above analysis, we can see that due to the angular momentum of the rotating black hole, the global behaviour of the generalized complexity at   early times can be more diverse than for static black holes. This is mainly due to the more complicated scalar function $F$ in \eqref{fonelg}. 
Although the effective potentials in the left diagrams of Figs. \ref{effpmp1}, \ref{effpmp2}, \ref{effpmp3}, and \ref{effpmp4} share similar configurations all have a generalized complexity that grows linearly at   late times, at   early times the phase behaviour of  generalized complexity varies for different specific parameters of the black holes and the coupling constant. 

Up to this point,  the transition times in the middle diagrams of Figs. \ref{effpmp1}, \ref{effpmp2}, \ref{effpmp3}, and \ref{effpmp4} were  determined only by the intersection points in the $C-\tau$ diagrams. One may wonder that whether there is an alternative way to determine   the phase transition time. The answer is yes and we will elaborate on this in what follows.

\subsection{Area law}\label{jf392j293}

\begin{figure}[t!]
	\centering
	\includegraphics[width=2.7in]{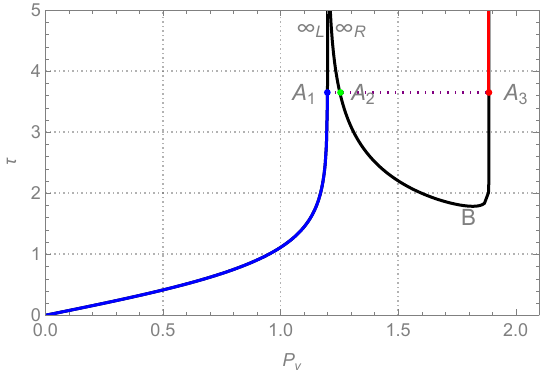}
	\caption{The variation of the boundary time with respect to the conserved momentum is examined. The boundary time undergoes a sudden shift from the blue branch to the red branch. \(A_1,\; A_2,\; A_3\) are the intersection points of the horizontal line, which signifies the transition time, with the solid curve. \(B\) denotes a point on the black curve. We have set $m=8/5, \;a=2/5,\; \ell=9/5,\; \lambda=1/10^{5}$  for the five-dimensional MP-AdS black hole. }\label{ajioee3}
\end{figure}

In what follows, we will show that an area law, similar to Maxwell's, can be used to determine the transition time of a phase transition for the generalized complexity. For simplicity, we will first concentrate on a simple case where there is only one transition time for the complexity, as shown in Fig. \ref{ajioee3}. According to the definition of the generalized volume complexity \eqref{codoneo}, the complexity will have a sudden transition from one phase (blue branch) to another phase (red branch). This phase transition is first-order, in the sense that the first-order derivative of the complexity with respect to time at the phase transition point is discontinuous. In Fig. \ref{ajioee3}, the real path of the boundary time curve has a discontinuous variation at the transition time, jumping from point \(A_1\) to point \(A_3\). Using \eqref{djio289}
we obtain
\begin{equation}
C_{A_1\tau_L} = \frac{\Omega_{3}}{\mathrm{G}_\mathrm{N} \ell} \int_{\tau_{A_1}}^{\tau_L} P_v (\tau) \, \mathrm{d}\tau 
\end{equation}
for the variation of the complexity from point \(A_1\) (with a boundary time \(\tau_{A_1}\)) to some point with a boundary time \(\tau_L\)  along the left branch of the curve in Fig. \ref{ajioee3}; when $\tau_L\to\infty$ (shown as $\infty_L$ in Fig. \ref{ajioee3}), we denote $C_{A_1\tau_L}$ as $C_{A_1\infty_L}$.  From a point on the right branch with a boundary time \(\tau_R\)  back to the point \(A_2\) with the boundary time \(\tau_{A_2} = \tau_{A_1}\), we have the variation of the complexity as
\begin{equation}
C_{\tau_R A_2} = \frac{\Omega_{3}}{\mathrm{G}_\mathrm{N} \ell} \int_{\tau_R}^{\tau_{A_2}} P_v (\tau) \, \mathrm{d}\tau\,.
\end{equation}
When $\tau_R\to\infty$ (denoted as $\infty_R$ in Fig. \ref{ajioee3}), we denote $C_{\tau_R A_2}$ as $C_{\infty_R A_2}$.
Similar to \cite{Belin:2022xmt,Jorstad:2022mls,Wang:2023eep}, at $\tau= \tau_L=\tau_R$, the net variation of the generalized volume complexity is equal to the difference between the complexity $C_{\tau_L}$ at boundary time $\tau_L$ and the complexity $C_{\tau_R}$ at boundary time $\tau_R$, i.e.,
\begin{equation}\label{jf93p2}
    \begin{aligned}
    \delta C(\tau_L= \tau_R) &=C_{\tau_L}- C_{\tau_R}\\
    &=\frac{\Omega_{3}}{\mathrm{G}_\mathrm{N} \ell}\left(\int_0^{\tau_L}P_{v}^L(\tau^\prime)\mathrm{d}\tau^\prime-\int_{0}^{\tau_R}P_v^R(\tau^\prime)\mathrm{d}\tau^\prime\right)\\&=\frac{2\Omega_{3}}{\mathrm{G}_\mathrm{N} \ell}\left(\int_{r_{\min }(\tau_L)}^{r_{\max }} \mathrm{d} r \frac{\mathfrak{a}(r)^2 h(r)^2 r^{2 d-6} f(r) g(r)}{\sqrt{\left(P_v^L\right)^2-U_0(r)}}\right.\\&\quad\quad\quad\quad\left.- \int_{r_{\min }(\tau_R)}^{r_{\max }} \mathrm{d} r \frac{\mathfrak{a}(r)^2 h(r)^2 r^{2 d-6} f(r) g(r)}{\sqrt{\left(P_v^R\right)^2-U_0(r)}}\right)\,,
    \end{aligned}
\end{equation}
where $P_v^L, P_v^R$ are the conserved momenta respectively corresponding to the boundary times $\tau_L, \tau_R$ according to \eqref{bdtau} and \eqref{urormin}, and they satisfy $P_v^L-P_\infty\leq 0^-$, $P_v^R-P_\infty\geq 0^+$.
When $\tau_L=\tau_R\to\infty$, we further get
\begin{equation}\label{jfi9328r}
    \begin{aligned}
    \delta C(\tau_L=\tau_R\to\infty) =&\frac{2\Omega_{3}}{\mathrm{G}_\mathrm{N} \ell}\left(\lim_{r^\prime\to r_f}\int_{r^\prime}^{r_{\max }} \mathrm{d} r \frac{\mathfrak{a}(r)^2 h(r)^2 r^{2 d-6} f(r) g(r)}{\sqrt{\left(P_{\infty}^L\right)^2-U_0(r)}}\right.\\&\quad\quad\quad\left.- \lim_{r^\prime\to r_R}\int_{r^\prime}^{r_{\max }} \mathrm{d} r \frac{\mathfrak{a}(r)^2 h(r)^2 r^{2 d-6} f(r) g(r)}{\sqrt{\left(P_{\infty}^R\right)^2-U_0(r)}}\right)\,,
    \end{aligned}
\end{equation}
where $r_f$ is the extremal point obtained by \eqref{conmu} and $r_R$ is the turning point satisfying \eqref{urormin} (cf. Fig. 8 in \cite{Belin:2022xmt}), and $P_{\infty}^L-P_{\infty}^R\to 0^-$.
From the right diagrams of Figs. \ref{effpmp1}, \ref{effpmp2}, \ref{effpmp3}, and \ref{effpmp4}, we can see that
\begin{equation}
    \delta C(\tau_L= \tau_R)>0\,;
\end{equation}
refer also to \cite{Belin:2022xmt} for a general explanation. Moreover, according to \eqref{vjiq398}, we have
\begin{equation}
    \delta C(\tau_L=\tau_R\to\infty)\to\mathrm{const.}>0\,.
\end{equation}

Then we have the total variation of the complexity \(C_{A_1A_2}\) along the path \(\overline{A_1\infty_L\infty_RA_2}\) as
\begin{equation}
C_{A_1A_2} = C_{A_1\infty_L} + C_{\infty_R A_2}-\delta C(\tau\to\infty)\,.
\end{equation}
The geometric meaning of the first two terms is the area of the region enclosed by the path \(\overline{A_1\infty_L\infty_RA_2A_1}\) in Fig. \ref{ajioee3}. The third term, explicitly expressed in \eqref{jfi9328r}, arises as there are two kinds of turning points $r_f,\, r_R$ sharing the same conserved momentum\footnote{We are grateful for the referee for pointing this out.}.

Similarly, going from point \(A_2\) 
to point \(A_3\) (with  boundary time \(\tau_{A_3}= \tau_{A_2}\)) via 
some point \(B\)  on the curve, we have  
\begin{equation}
C_{A_2 A_3} = \frac{\Omega_{3}}{\mathrm{G}_\mathrm{N} \ell} \int_{\tau_{A_2}}^{\tau_{A_3}} P_v (\tau) \, \mathrm{d}\tau\,.
\end{equation}
The geometric meaning of this is the area of the region enclosed by the path \(\overline{A_2BA_3A_2}\), shown in Fig. \ref{ajioee3}. As the phase transition of the complexity is first-order, meaning that the variation of the complexity is continuous at the transition point, we know that the complexity at point \(A_1\) is the same as that at point \(A_3\). This means that
\begin{equation}\label{dijopa23}
C_{A_1A_2} + C_{A_2A_3} = 0\,.
\end{equation}
Geometrically, we can interpret this by using the areas of the corresponding  two regions. To this end, we first denote the area of the region enclosed by the path \(\overline{A_1\infty_L\infty_RA_2A_1}\) as \(\mathcal{S}_{A_1A_2} \equiv |C_{A_1\infty} + C_{\infty A_2}|\). Since there is a divergence of the boundary time $\tau$ in this region, we should modify the area by a constant, yielding 
\begin{equation}
\tilde{\mathcal{S}}_{A_1A_2}=\mathcal{S}_{A_1A_2}-\delta C(\tau\to\infty)
\label{newarea}
\end{equation}
as the modification $\tilde{\mathcal{S}}_{A_1A_2}$ for the  area. Denoting the area of the region enclosed by the path \(\overline{A_2BA_3A_2}\) as \(\mathcal{S}_{A_2A_3} \equiv |C_{A_2A_3}|\). As a result, we have an  area law  
\begin{equation}\label{adj8439}
\tilde{\mathcal{S}}_{A_1A_2} =\mathcal{S}_{A_2A_3} 
\end{equation}
for the first-order phase transition of the generalized complexity from the blue branch to the red branch in Fig. \ref{ajioee3}.

In other words,  if we can find a horizontal line demarking two regions
that satisfy  \eqref{adj8439} fulfilled,  there will be a first-order phase transition for the complexity. If such a horizontal line does not exist, then there is no phase transition for the complexity, and only the phase represented by the red curve exists.

To generalize this area law to more involved cases, let us analyze the type (A), (B), and (C) phase transitions shown in the previous section \ref{sec:gvcse}. For simplicity, we will not draw diagrams again. We use \(\mathcal{S}_{ij}\) to denote the area of a region enclosed by a horizontal line connecting two points \(X_i\) and \(X_j\) and the variation curve of the boundary time \(\tau\) with respect to the conserved momentum \(P_v\) in the middle diagrams of Figs. \ref{effpmp1}, \ref{effpmp2}, and \ref{effpmp3}. For the type (A) phase transition, we can denote the intersection point set of the horizontal line in the middle diagram of Fig. \ref{effpmp1} from left to right as \(\left\{A_i \mid i = 1, 2, \cdots, 5\right\}\). We then have
\begin{equation}\label{jf923}
\tilde{\mathcal{S}}_{A_1 A_2} + \tilde{\mathcal{S}}_{A_3 A_4} = \mathcal{S}_{A_2 A_3} + \mathcal{S}_{A_4 A_5}\,,
\end{equation}
where, as in \eqref{newarea},  the original areas $\mathcal{S}_{A_1 A_2}$ and $\mathcal{S}_{A_3 A_4}$ on the left side are replaced by the modified areas $\tilde{\mathcal{S}}_{A_1 A_2}=\mathcal{S}_{A_1 A_2}-\delta C_1(\tau\to \infty)$ and $\tilde{\mathcal{S}}_{A_3 A_4}=\mathcal{S}_{A_3 A_4}-\delta C_2(\tau\to \infty)$. $\delta C_1(\tau\to \infty)$ and $\delta C_2(\tau\to \infty)$ arise due to the divergences of the boundary time in the corresponding regions. In what follows, we will denote the modified areas by $\tilde{\mathcal{S}}$ and will not emphasize this again.

For the type (B) phase transition, we can denote the intersection point set of the lower horizontal line in the middle diagram of Fig. \ref{effpmp2} from left to right as \(\left\{A_i \mid i = 1, 2, 3\right\}\) and the intersection point set of the higher horizontal line from left to right as \(\left\{A_i^\prime \mid i = 3, 4, 5\right\}\); we  then  have
\begin{equation}\label{fj2938}
\tilde{\mathcal{S}}_{A_1 A_2} = \mathcal{S}_{A_2 A_3}\,, \quad \tilde{\mathcal{S}}_{A_3^\prime A_4^\prime} = \mathcal{S}_{A_4^\prime A_5^\prime}\,.
\end{equation}
For the type (C) phase transition, we have, likewise,
\begin{equation}
\tilde{\mathcal{S}}_{A_3 A_4} = \mathcal{S}_{A_4 A_5}\,.
\end{equation}

\begin{figure}[t!]
    \centering
    \includegraphics[width=0.5\linewidth]{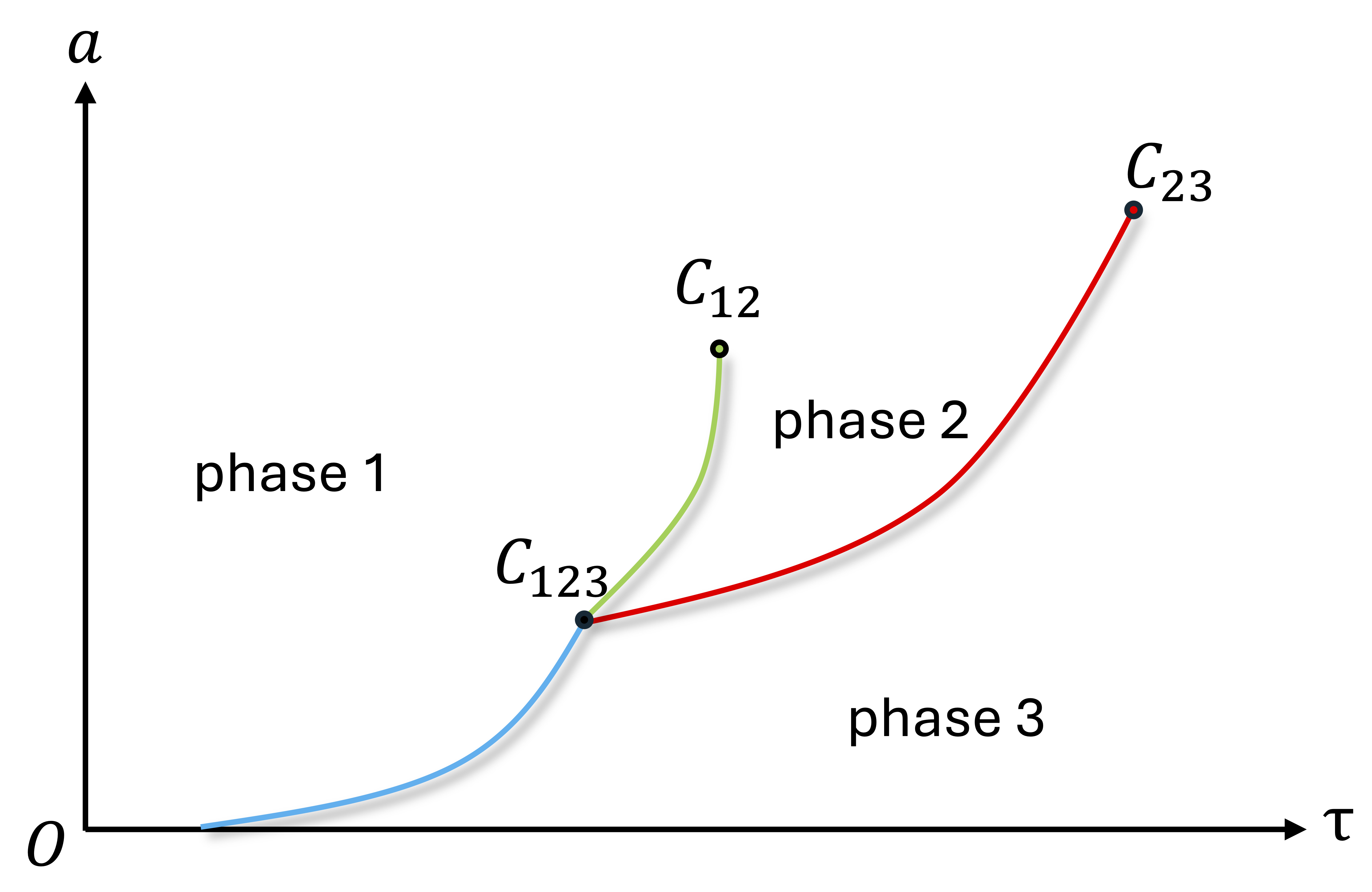}
    \caption{A typical diagram of the phase transition for the generalized holographic complexity  in a foliation with constant $m,\; \ell,\; \lambda$. $C_{12},\,C_{23}$ are critical points of the second-order phase transitions from phases 1 to 2 and from phases 2 to 3, respectively. $C_{123}$ is a multi-critical point where two first-order phase transitions between phase 1 and 2 and between phase 2 and 3 share the same parameter configuration. The green, red, and blue curves represent the ``coexistence curves'' bewtween phases 1 and 2, 2 and 3, and 1 and 3, respectively.}
    \label{fj938z}
\end{figure}

The area law we observe for   phase transitions of   generalized holographic complexity is similar to   Maxwell's equal area law. However, it is not  applicable to a thermodynamic process but rather  deals with the chronological variation of the complexity. This area law indicates that \(\int P_v(\tau) \, d\tau\) converges over any integral region, even one that contains a divergent \(\tau\) at critical momentum \(P_v\). 

 For a type (B) phase transition of the generalized complexity, when the purple line in the right diagram of the Fig. \ref{effpmp2} vanishes, the two first-order phase transition points merge into a multi-critical point. Under this condition  \eqref{jf923} and \eqref{fj2938} are fulfilled at the same time, yielding 
\begin{align}\label{jfi9283}
\tilde{\mathcal{S}}_{A_1 A_2}  =\mathcal{S}_{A_2 A_3} \,,\quad\tilde{\mathcal{S}}_{A_3 A_4} = \mathcal{S}_{A_4 A_5}\,.
\end{align}
In this case, the type (B) phase transition happens to reduce to the type (A) phase transition. It is worth noting that the condition  \eqref{jf923} does not necessarily imply \eqref{jfi9283};  the type (A) phase transition does not necessarily admit a multi-critical point. 

By using the  area law, we can infer that there can be second-order phase transitions of generalized complexity. For example, considering the phase transition shown in Fig. \ref{effpmp2}, a second-order phase transition will take place between phases 1 and 2
if 
\begin{equation}\label{afj3939}
\tilde{\mathcal{S}}_{A_1A_2} = \mathcal{S}_{A_2A_3}=0\,,
\end{equation}
which necessitates a specific choice of   parameters (such as the coupling parameter). This is in fact the case where $r_M$ and $r_R$ merge in the left diagram of Fig. \ref{effpmp2}. In this situation  we have $\mathcal{S}_{A_1A_2}\to 0$ as well as the corresponding area modification $\delta C\to 0$. As with second-order thermodynamic phase transitions for  black holes \cite{Kubiznak:2012wp},  the two complexity phases cannot be distinguished.
The complexity and its first derivative   with respect to the boundary time will be continuous, but the second derivative will not be continuous.  Moreover, under certain conditions we can have one other second-order phase transition between phases 2 and 3 if 
\begin{equation}\label{fj2d9mds}
    \tilde{\mathcal{S}}_{A_3 A_4} = \mathcal{S}_{A_4 A_5}=0\; ,
\end{equation}
  similar to \eqref{afj3939}.

The scenario here is quite similar to the multi-critical phase transitions discovered for the thermodynamics of nonlinear electromagnetic black holes \cite{Tavakoli:2022kmo,Fang:2022rsb}, multiply rotating black holes \cite{Wu:2022bdk}, and black holes in higher-curvature gravitational theories \cite{Wu:2022plw,Lu:2023hgu}. In light of the above analysis, we qualitatively draw a phase diagram showing the phase structures of the generalized holographic complexity in Fig. \ref{fj938z}. In units of $\ell$, we have free parameters $a,\; m, \;\lambda$. With a finely tuned parameter configuration, we can have the foliation shown in Fig. \ref{fj938z}. By this diagram, we can easily identify the four types of phase structures shown Figs. \ref{effpmp1}, \ref{effpmp2}, \ref{effpmp3}, and \ref{effpmp4}. Additionally, the multi-critical point $C_{123}$ corresponds to the condition \eqref{jfi9283}, and the critical points $C_{12}$ and $C_{23}$ satisfy \eqref{afj3939} and \eqref{fj2d9mds}, respectively. From the diagram, we can see that there will only be a phase transition for the complexity between phases 1 and 3 in the $a\to 0$ limit, i.e., the Schwarzschild-AdS limit; this is consistent with the reslult shown in \cite{Belin:2022xmt}.

We close this section by noting that in the above analysis we used $\delta C\left(\tau_L=\tau_R \rightarrow \infty\right)$. One can alternatively obtain an area law by applying $\delta C\left(\tau_L=\tau_R\right)$ in \eqref{jf93p2}, but  all derived results would be the same. The Maxwell-like modified area law is a direct consequence of  the time derivative of the generalized complexity being  absolutely determined by the conserved momentum through \eqref{djio289}.

\section{Generalized volume complexity of formation}\label{fej3i9ejfpq}

\subsection{MP-AdS black hole case}

Unlike   spherically symmetric black holes \cite{Chapman:2016hwi, Carmi:2017jqz}, it is the thermodynamic volume rather than the entropy that determines the behaviour of the volume complexity of formation for large rotating black holes
 \cite{AlBalushi:2020rqe, AlBalushi:2020heq}. 
In this section, we will study the complexity of formation in the generalized volume complexity framework for the odd-dimensional MP-AdS black holes.

The former proposal for the volume complexity of formation for black holes is defined by subtracting the volume complexity of   vacuum AdS from that of the black hole at the boundary time \(\tau=0\) \cite{Chapman:2016hwi}. With the definition of generalized complexity \eqref{codoneo}, we can introduce the concept of the generalized volume complexity of formation, defined by
\begin{equation}\label{ji323498}
\Delta C = C_{\tau=0}\left(\Sigma_{\mathrm{CFT}}\right) - C^{\mathrm{AdS}}_{\tau=0}\left(\Sigma_{\mathrm{CFT}}\right)\,,
\end{equation}
where the first term on the right-hand side of \eqref{ji323498} is determined by calculating the generalized volume complexity at the boundary time \(\tau=0\) via \eqref{vmax2}. Note that at this moment, we have \(r_{\min} = r_{+}\) and \(P_v^2 = 0\). The second term on the right-hand side of \eqref{ji323498} is the generalized volume complexity 
\begin{equation}
\begin{aligned}
C_{\tau=0}^{\mathrm{AdS}} &= \max_{\partial \Sigma = \Sigma_{\mathrm{CFT}}} \frac{\mathcal{V}_{\mathrm{AdS}}(\tau=0,\,m=0)}{\mathrm{G}_\mathrm{N} \ell} \\
&= \max_{\partial \Sigma(\tau=0) = \Sigma_{\mathrm{CFT}}} \frac{2\Omega_{d-2}}{\mathrm{G}_\mathrm{N} \ell} \int_0^{r_{\max}^{\mathrm{AdS}}} \mathrm{d}r \frac{r^{d-2}}{\sqrt{1 + r^2 / \ell^2}} F^{\mathrm{AdS}}\left(g_{\mu \nu}, \mathcal{R}_{\mu \nu \rho \sigma}, \nabla_\mu\right)
\end{aligned}
\end{equation}
of   the  vacuum AdS, where the scalar function $F^{\mathrm{AdS}}\left(g_{\mu \nu}, \mathcal{R}_{\mu \nu \rho \sigma}, \nabla_\mu\right)$  is determined by the curvature invariants of the vacuum AdS  geometry, obtained from the gravitational observable \(F\left(g_{\mu \nu}, \mathcal{R}_{\mu \nu \rho \sigma}, \nabla_\mu\right)\) in \eqref{codoneo} in the vacuum limit.  

For   vacuum AdS\(_d\), the extremal value of the generalized volume functional reduces to
\begin{equation}
\mathcal{V}_{\mathrm{AdS}} = 2\Omega_{d-2} \int_0^{r_{\max}} \mathrm{d}r \frac{r^{d-2} \mathfrak{a}(m=0)}{\sqrt{1 + r^2 / \ell^2}}\,.
\end{equation}
Thus, the generalized volume complexity of formation can be defined as
\begin{equation}\label{j3i98}
\begin{aligned}
  \mathrm{G}_\mathrm{N} \ell  \Delta C &= \mathcal{V}_{\max }(\tau=0) - \mathcal{V}_{\mathrm{AdS}}(\tau=0,\,m=0) \\
  &= 2\Omega_{d-2} \int_{r_{+}}^{r_{\max}} \mathrm{d}r \frac{\mathfrak{a}(r)^2 h(r)^2 r^{2d-6} f(r) g(r)}{\sqrt{-U_0(r)}} - 2\Omega_{d-2} \int_0^{r_{\max}} \mathrm{d}r \frac{2 (120 \lambda + 1) r^3}{\sqrt{1 + r^2 / \ell^2}}\,,
\end{aligned} 
\end{equation}
where we have used the conditions \(r_{\min} = r_{+}\) and \(P_v^2 = 0\) at \(\tau=0\). We have also chosen in \eqref{j3i98}  the Gauss-Bonnet invariant \eqref{fji933498} to be the curvature invariant, yielding  the explicit form of \(\mathcal{V}_{\mathrm{AdS}}(\tau=0,\,m=0)\). For the black hole, we use the UV cut-off \cite{AlBalushi:2020heq}
\begin{equation}
   r_{\max} = \rho - \frac{\ell^2}{4 \rho} + \frac{\ell^2 M \Xi}{(2 N + 2) \rho^{2 N + 1}} + \mathcal{O}\left(\rho^{-(2 N + 3)}\right)\,,
\end{equation}
whereas for the vacuum AdS background, we take
\begin{equation}
r_{\max}^{\mathrm{AdS}} = \rho - \frac{\ell^2}{4 \rho}\,,
\end{equation}
where \(\rho = \ell^2 / \delta\) with \(\delta \to 0\).

We now investigate  the scaling law of the generalized volume complexity of formation with respect to thermodynamic volume for the large rotating black hole.   The thermodynamic entropy and the volume of the rotating black hole are
\begin{equation}
S = \frac{\Omega_{2N+1} h\left(r_{+}\right) r_{+}^{2N}}{4 \mathrm{G}_\mathrm{N}}\,,
\end{equation}
\begin{equation}
V = \frac{r_{+}^{2(N+1)} \Omega_{2N+1}}{2(N+1)} + \frac{4 \pi a J}{(2N+1)(N+1)}\,,
\end{equation}
where the angular momentum \(J\) of the black hole is 
\begin{equation}
J = \frac{\Omega_{2N+1}}{4 \pi \mathrm{G}_\mathrm{N}} (N+1) m a\,.
\end{equation}
In the near extermal limit, the entropy and the volume of the black hole 
respect different scaling relations
\begin{equation}
\lim_{r_{-}/r_{+} \rightarrow 1} S \sim \left(\frac{r_{+}}{\ell}\right)^{2N+2}, \quad \lim_{r_{-}/r_{+} \rightarrow 1} V \sim \left(\frac{r_{+}}{\ell}\right)^{2N+4}
\end{equation}
with respect to $r_+$.

The volume complexity of formation for the rotating black hole \eqref{mpadsmet} was shown to scale like \cite{AlBalushi:2020rqe, AlBalushi:2020heq}
\begin{equation}\label{ji932}
\Delta C \sim V^{(d-2) /(d-1)}
\end{equation}
for large black holes, a relation that becomes most distinct from scaling with entropy as extremality is approached.
Here we investigate the scaling behaviour of
$\Delta C$ in the context of  generalized volume complexity of formation defined by \eqref{ji323498} and specified in \eqref{j3i98}.

 To do so, we consider the ratio
\begin{equation}\label{jfio3q49}
\bar{\mathcal{R}} = \frac{\Delta C}{\left(r_+/\ell\right)^\beta}
\end{equation}
in the   near-extremal limit \(r_-/r_+ \to 1\) for a large black hole with \(r_+/\ell \gg 1\).  If the scaling relation \eqref{ji932} holds
in the generalized case, then 
 \(\bar{\mathcal{R}}\) will be constant 
 over a broad range of coupling parameters
 $\lambda$ provided $\beta = (d-2)(d+1)/(d-1)$. We can check this numerically by computing the slope
 \begin{equation}\label{jfo33489}
\mathcal{S} \sim \frac{\bar{\mathcal{R}} \left(r_+/\ell + \delta\right) - \bar{\mathcal{R}} \left(r_+/\ell\right)}{\delta}
\end{equation}
over a range of values of $r_+$ for different values of $\lambda$,
where \(\delta\) denotes the interval between two sampling points on the curve.
 
\begin{figure}[t!]
	\centering
	\includegraphics[width=5.7in]{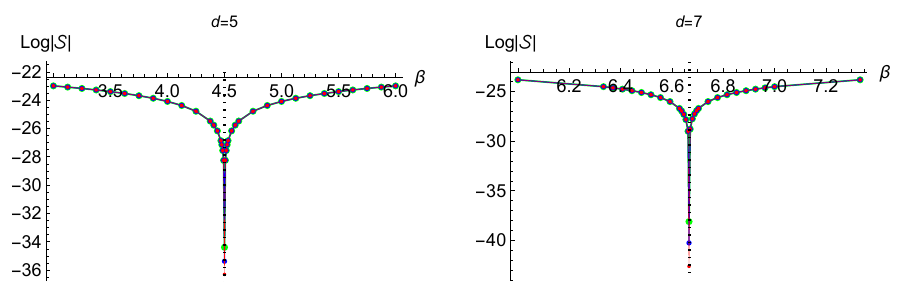}
	\caption{The variation of the slope of \(\mathcal{R}\) with respect to the dimensionless scaling parameter \(\beta\) is analyzed. We have set \(r_-/r_+ = 1 - 10^{-10}\), \(\ell = 1\), and \(\lambda = 1/100, 100, 0\) for the blue, red, and green points (curves), respectively. The slope of the curve is determined by choosing two points at \(r_+/\ell = 10^{10}\) and \(2 \times 10^{10}\). The UV cut-off is set to \(\delta = 10^{-24}\). A smaller UV cut-off is needed if we consider larger black holes. In the left diagram, \(\mathcal{S}\) attains minimal values at \(\beta = 4.500\) in all cases for \(d = 5\), while in the right diagram, \(\mathcal{S}\) attains minimal values at \(\beta = 6.667\) in all cases for \(d = 7\). We note that the result will be the same for negative (small) \(\lambda\). }\label{jef938}
\end{figure}

The results are shown in 
 Fig. \ref{jef938} for three different values of $\lambda$ for each of
$d=5,7$. We see that \(\mathcal{S}\) vanishes at $\beta = (d-2)(d+1)/(d-1)$  for all values of $\lambda$. In other words, the scaling behaviour of the generalized volume complexity still respects \eqref{ji932}, commensurate with the $\lambda=0$ case \cite{AlBalushi:2020rqe, AlBalushi:2020heq}. 
 One can check that if we choose a different scalar function from \eqref{fonelg} and the generalized complexity of formation can be well-defined (see below), the scaling behaviour will be the same.

These results indicate the universality of the scaling behaviour of the generalized volume complexity of formation. In what follow, we suppose that the scalar function $F>0$ in \eqref{codoneo} is defined by a form of 
\begin{equation}\label{fjpi392}
    F=1+\lambda \times \mathrm{curvature\,\, invariant\,\, terms}\,,
\end{equation}
so that in the vanishing coupling case, we have the original definition for the CV complexity. We observe that 
\begin{equation}\label{gj309}
\lim_{r \to \infty} F =1+ \text{const.} < \infty
\end{equation}
in the large \(r\) limit for all cases of scalar functions that are
 polynomials in the curvature invariants.\footnote{For instance, for the curvature invariants for the five-dimensional MP-AdS black hole, we can check that
$\lim_{r \to \infty} \mathcal{R}_{\mu\nu\rho\sigma} \mathcal{R}^{\mu\nu\rho\sigma} = 40/\ell^2 < \infty\,.$} Indeed, this is true for the MP-AdS black holes as vacuum solutions: such curvature invariants will be finite at spatial infinity. 

 Are there any versions of generalized complexity that are different from the scaling relation \eqref{ji932}?
If a function $F$ in \eqref{codoneo} were chosen so that
\(\lim_{r \to \infty} F \to \infty\), this might be possible. For example, noting that  
\begin{equation}
\lim_{r \to \infty} \mathcal{C}_{\mu\nu\rho\sigma} \mathcal{C}^{\mu\nu\rho\sigma} = 0\,,
\end{equation}
a scalar function such as \(F = 1 + \lambda \ell^{-4} \mathcal{C}^{-2}\) could be constructed.   However, it is evident from the result   in Appendix \ref{app:cio} that \(\mathcal{C}^{-2} \propto 1/m^2\), which makes it unclear how to define the generalized volume complexity of formation, as \(C_{\tau=0}^{\mathrm{AdS}}\) will be ill-defined in the vanishing mass limit.  As a result, if the scalar function satisfies the conditions 
\begin{equation}\label{fji39pzl}
    F(m\to 0)<\infty\,,
\end{equation}
such that the generalized complexity of formation can be well-defined, the scaling relation of the generalized volume complexity of formation will be aligned with \eqref{ji932}.

On the other hand, in the static limit with \( r_-/r_+ \to 0 \), it was argued that the scaling behaviour of the volume complexity of formation for the large non-rotating black hole is controlled, instead, by the entropy of the black hole \cite{AlBalushi:2020rqe, AlBalushi:2020heq}. Specifically, the scaling behaviour is
\begin{equation}\label{fpoi3e8}
\lim_{r_-/r_+ \to 0} \frac{\Delta C}{S} \to \tilde{k}_d\,,
\end{equation}
where \(\tilde{k}_d\) is a dimension-dependent constant. This result is the same as the one given by \cite{Chapman:2016hwi, Carmi:2017jqz} for the charged RN-AdS black holes.

\begin{figure}[t!]
	\centering
	\includegraphics[width=3.4in]{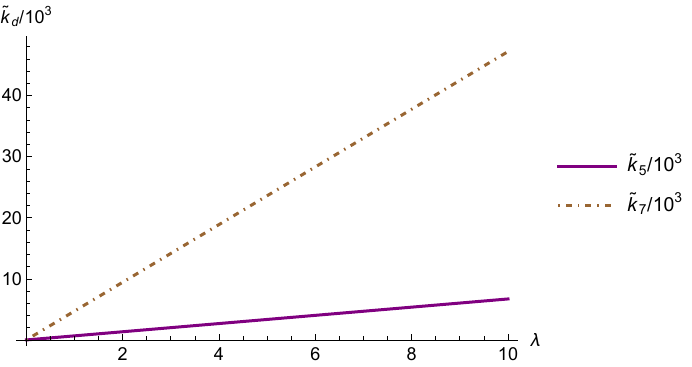}
	\caption{The variation of \( \tilde{k}_5\) with respect to the coupling parameter $\lambda$ for the five-dimensional MP-AdS black hole in the static limit is analyzed. We choose the parameters as \(\ell = 1\), \(r_-/r_+ = 10^{-18}\), \(r_+/\ell = 10^{10}\), and \(\delta = 10^{-30}\). The result here is consistent with the one obtained in \cite{Chapman:2016hwi} for the planar Schwarzschild-AdS black hole.}\label{djklaj3}
\end{figure}

Will the scaling behaviour \eqref{fpoi3e8}   be preserved for  generalized volume complexity of formation for   large rotating black holes?  We again employ the scalar function \eqref{fonelg} to check this for the five- and seven-dimensional MP-AdS black holes in their static limit (which are five- and seven-dimensional Schwarzschild-AdS black holes). We vary the coupling constant and find that \eqref{fpoi3e8} is always respected, though the constant \(\tilde{k}_d\) varies with respect to \(\lambda\), as shown in Fig. \ref{djklaj3}. These results indicate that in the static limit of the MP-AdS black hole, the generalized volume complexity of formation still scales with the entropy.

\subsection{Comparing with charged black hole case}

In the preceding section, we have checked that the scaling behaviour of the generalized volume complexity of formation for the large near-extremal rotating black hole is controlled by the volume of the black hole, independent of the values of the scalar function. We may ask whether the scaling behaviour of the charged RN-AdS black hole, which is controlled by the entropy of the black hole, will deviate from the original result discovered in \cite{Chapman:2016hwi, Carmi:2017jqz} if the generalized volume complexity of formation for the RN-AdS black hole is considered. We will study this in what follows.

For simplicity, we directly use the result of the generalized volume complexity for the RN-AdS black hole \cite{Jorstad:2023kmq, Wang:2023eep}
\begin{equation}
C_{\text{gen}} = \frac{2\Omega_{d-2} \ell^{d-3}}{\mathrm{G}_\mathrm{N}} \int_{r_{\text{min}}}^{r_{\text{max}}} \mathrm{d}r \frac{(r / \ell)^{2d-4} \mathfrak{a}^2(r)}{\sqrt{\mathfrak{P}_v^2 - \mathcal{U}_0(r)}}\,,
\end{equation}
where
\begin{equation}
\mathcal{U}_0(r) = -\mathfrak{f}(r) \mathfrak{a}^2(r) \left(\frac{r}{\ell}\right)^{2(d-2)}\,,
\end{equation}
\begin{equation}\label{fjo29}
\mathfrak{f}(r) = k - \frac{16 \pi M}{(d-2) \Omega_{d-2}} \cdot \frac{1}{r^{d-3}} + \frac{32 \pi^2 Q^2}{\left(d^2 - 5d + 6\right) \Omega_{d-2}^2} \cdot \frac{1}{r^{2(d-3)}} + \frac{r^2}{\ell^2}\,,
\end{equation}
and the scalar function \(\mathfrak{a}(r)\) is obtained by evaluating \(F\) (see \eqref{codoneo}) in the RN-AdS background. Here, \(M\) and \(Q\) are individually  the mass and electric charge of the black hole. \(k = 0, 1, -1\) are for planar, spherical, and hyperbolic horizons, respectively. \(\mathfrak{P}_v\) is the conserved momentum. For efficiency of calculation, we choose \(k = 0\) in what follows. The event horizon \(r_+\) of the RN-AdS black hole can be obtained via \(\mathfrak{f}(r) = 0\).

There are typically two kinds of asymptotic properties for the scalar function \(F\). For the first kind, \eqref{gj309} is satisfied.
The curvature invariant terms in \eqref{fjpi392} include \(C^2\), \(\mathcal{G}\), or other combinations of the curvature invariants \(\mathcal{R}^2\), \(\mathcal{R}_{\mu\nu}\mathcal{R}^{\mu\nu}\), \(\mathcal{R}_{\mu\nu\rho\sigma}\mathcal{R}^{\mu\nu\rho\sigma}\). The other kind of function becomes divergent at spatial infinity, i.e.,
\begin{equation}\label{jf39q8qnz}
  \lim_{r\to\infty}F\to \infty\,,  
\end{equation}
such as \(\left(\mathcal{R}_{\mu\nu} \mathcal{R}^{\mu\nu} - \mathcal{R}^2/4\right)^{-1/8}\). We find that these two kinds of scalar functions can lead to quite different scaling behaviours of the generalized complexity of formation for the RN-AdS black holes.

According to \eqref{ji323498}, the explicit form of the generalized volume complexity of formation for the RN-AdS black hole is
\begin{equation}\label{j9ie238}
\Delta C = \frac{2 \Omega_{d-2} \ell^{d-3}}{\mathrm{G}_\mathrm{N}} \int_{r_{+}}^{r_{\text{max}}} \mathrm{d}r \frac{(r / \ell)^{2d-4} \mathfrak{a}^2(r)}{\sqrt{-\mathcal{U}_0(r)}} - \frac{2 \Omega_{d-2} \ell^{d-3}}{\mathrm{G}_\mathrm{N}} \int_{r_{\text{min}}}^{r_{\text{max}}} \mathrm{d}r \frac{(r / \ell)^{d-2} \mathfrak{a}(r)}{\sqrt{1 + r^2 / \ell^2}}\,,
\end{equation}
where the definition of the function \(F\) is different from the one in \eqref{fonelg}. Here, according to our mentioned two different categories of the specific values of the function \(\mathfrak{a}(r)\), we choose it to be
\begin{equation}\label{dji2993}
F_1(r) = 1 + \lambda \ell^4 C^2
\end{equation}
for the first kind of scalar function, where the Weyl scalar reads
\begin{equation}\label{j293}
C^2 = \frac{48\left(Q^2 - M r\right)^2}{r^8}\,.
\end{equation}
We choose 
\begin{equation}\label{ji9e238}
F_2(r) = 1 + \frac{\lambda}{\ell^{4x}} \left(R_{\mu\nu} R^{\mu\nu} - \frac{R^2}{4}\right)^{-x} = 1 + \frac{4\lambda r^{8x}}{(Q \ell)^{4x}}
\end{equation}
for the second kind of scalar function, where $x$ is a dimensionless parameter. 
For clarity, we will denote the generalized complexity of formation corresponding to the first and second kinds of functionals as \(\Delta C_1\) and \(\Delta C_2\), respectively.

\subsubsection{Case 1: the same scaling}

\begin{figure}[t!]
	\centering
	\includegraphics[width=5.7in]{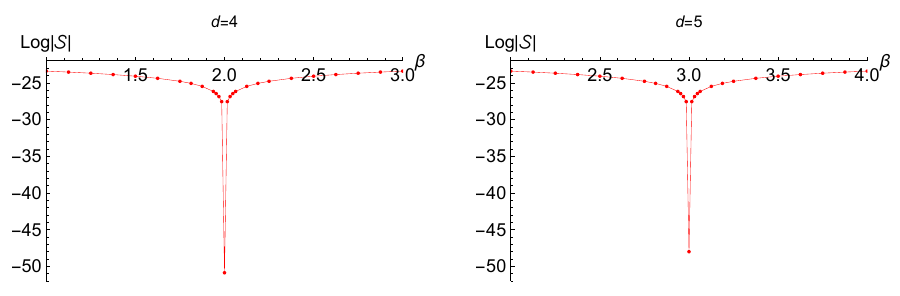}
	\caption{The variation of the slope of \(\bar{\mathcal{R}}\) with respect to the  scaling dimension \(\beta\) is analyzed for the RN-AdS black hole. We have set \(r_-/r_+ = 1 - 10^{-10}\) and \(\ell = 1\). We let \(\lambda = 10\) and \(1/10\) for the \(d = 4\) and \(d = 5\) cases, respectively (we have checked that the specific values of \(\lambda\) do not affect the scaling dimensions). Additionally, we have taken \(x = -1/4\) for both \(d = 4\) and \(d = 5\) cases. The slope of the curve is determined by two points at \(r_+/\ell = 10^{10}\) and \(2 \times 10^{10}\). The UV cut-off is set to \(r_{\mathrm{max}} = 10^{20}\). }\label{j398dweio}
\end{figure}

 For the choice \eqref{dji2993}
the results are shown in Fig. \ref{j398dweio}, using the methods described above.
We find that the scaling behaviour of the generalized volume complexity of formation is the same as the scaling behaviour of the original complexity of formation; both scale with the entropy $ S \sim r_+^{d-2}$
of the RN-AdS black hole, commensurate with the scaling behaviour of the volume complexity of formation for the RN-AdS  black hole \cite{AlBalushi:2020heq}.

\begin{figure}[t!]
	\centering
	\includegraphics[width=5.7in]{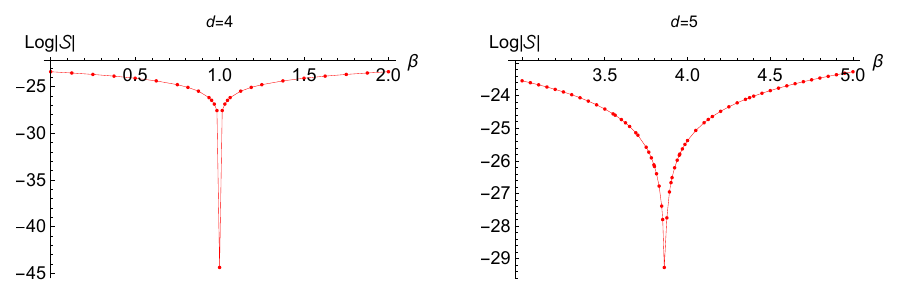}
	\caption{The variation of the slope of \(\bar{\mathcal{R}}\) with respect to the dimensionless scaling dimension \(\beta\) for the RN-AdS  black hole  
 for the choice \eqref{ji9e238}
with \(x = 1/4\) for both \(d = 4\) and \(d = 5\).
We have set \(r_-/r_+ = 1 - 10^{-10}\) and \(\ell = 1\). We let \(\lambda = 10\) and \(1/10\) for the \(d = 4\) and \(d = 5\) cases, respectively (we have checked that the specific values of \(\lambda\) do not affect the scaling dimensions).  The slope of the curve is determined by choosing two points at \(r_+/\ell = 10^{10}\) and \(2 \times 10^{10}\). The UV cut-off is set to \(r_{\mathrm{max}} = 10^{20}\). }\label{j398dweio2}
\end{figure}

\subsubsection{Case 2: the deviation of the scaling law}

 Choosing \eqref{ji9e238}, we 
obtain   the scaling behaviour   shown in Fig. \ref{j398dweio2}.
It is evident that the scaling behaviour of this choice for generalized volume complexity of formation is {\it not} controlled by the entropy of the charged black hole.  For example, when $d=4$, we find that the scaling dimensions of the volume complexity of formation varies with respect to the parameter \(x\) by the relation
\begin{equation}
\beta = -8x + 3\,.
\end{equation}
This relation holds only for \(x > 0\). For \(x = 0\), we have \(\beta = 2\).

These discoveries indicate that if the scalar function that defines  generalized holographic complexity is asymptotic to a constant quantity at spatial infinity, then both the scaling laws of the generalized volume complexity of formation for the rotating MP-AdS black holes and  spherically symmetric black holes will be the same as those for the original complexity of formation under the CV proposal. However, if the  scalar function that  defines   generalized holographic complexity becomes infinite as   spatial infinity is approached, the 
generalized volume complexity of formation cannot be defined for the MP-AdS black holes and the scaling law for  spherically symmetric black holes will deviate from that of the original complexity of formation.

\section{Codimension-zero observables of MP-AdS black holes}\label{jfoi39}

Thus far we have studied the generalized volume complexity and its formation for the MP-AdS black holes. The generalized volume complexity is a codimension-one observable. In the ``complexity equals anything'' proposal, there is another observable that is codimension-zero. Here we show   how this observable can be constructed for the MP-AdS black hole and show its similarities with the codimension-one observable. Due to the similarities they share, we will not further study the phase transitions and the complexity scaling law, which can be easily calculated.

\begin{figure}
    \centering
    \includegraphics[width=1.8in]{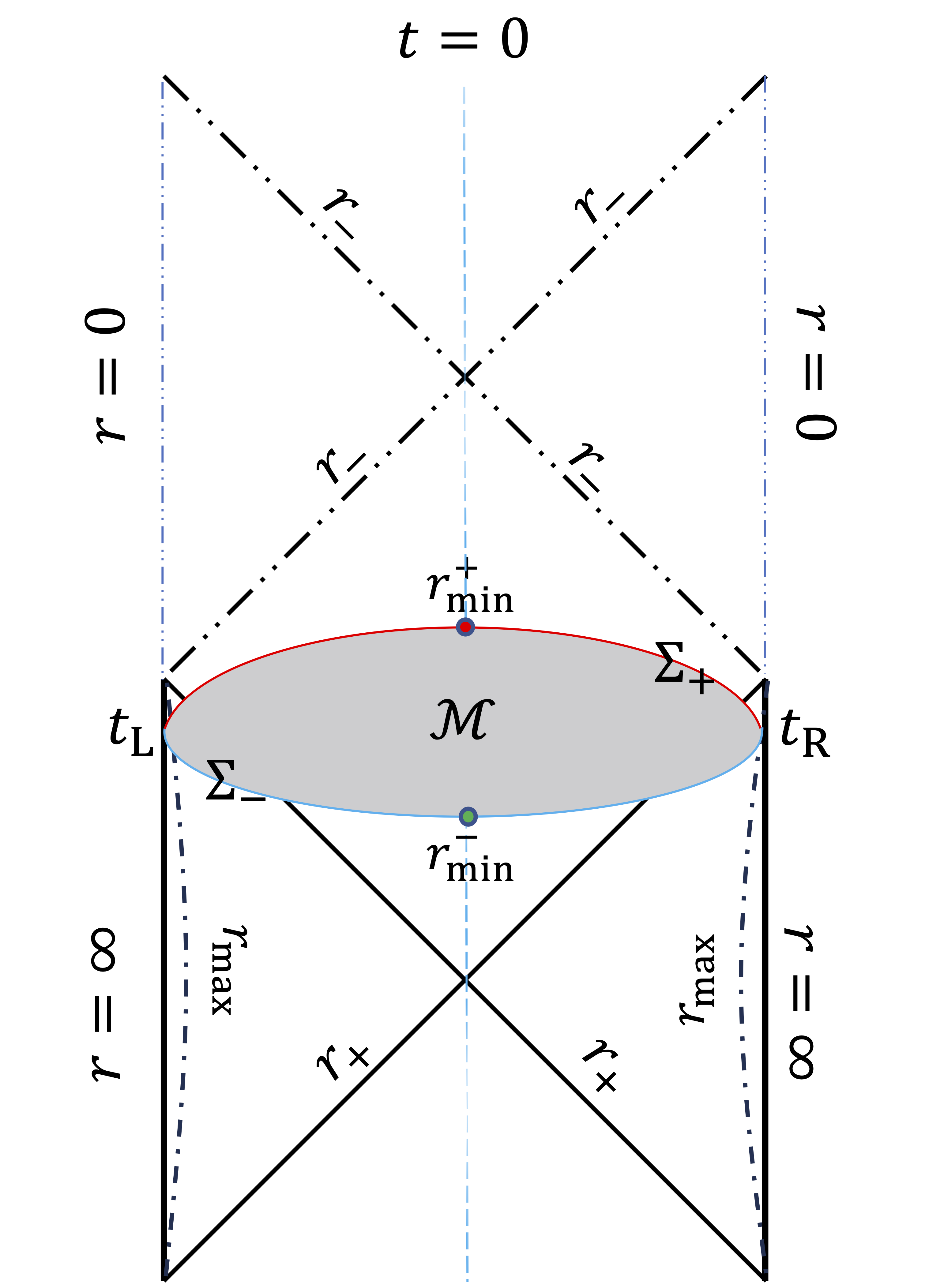}
    \caption{In the Penrose diagram of the MP-AdS black hole, $\mathcal{M}$ stands for the codimension-zero region, which is bounded by two hypersurfaces $\Sigma_{\pm}$. $r_{\min}^{\pm}$ are the turning points of the psudoparticles regarding to the two boundaries.}
    \label{fji399j2}
\end{figure}

Codimension-zero generalized complexity  $\mathcal{C}_\mathrm{0}(\tau)$ can be defined as (see Fig. \ref{fji399j2}) \cite{Belin:2022xmt}
\begin{equation}
\begin{aligned}
C_\mathrm{0}\left(\Sigma_{\mathrm{CFT}}\right)= & \max _{\partial \Sigma_{ \pm}=\Sigma_{\mathrm{CFT}}}\left[\frac{1}{\mathrm{G}_\mathrm{N} \ell^2} \int_{\mathcal{M}_{G, F_{ \pm}}} \mathrm{d}^{d} x \sqrt{-g} G\left(g_{\mu \nu}\right)\right. \\
& \left.+\frac{1}{\mathrm{G}_\mathrm{N} \ell} \int_{\Sigma_{+}} \mathrm{d}^{d-1} \sigma \sqrt{\tilde{h}} F_{+}\left(g_{\mu \nu} ; X_{+}^\mu\right)+\frac{1}{\mathrm{G}_\mathrm{N} \ell} \int_{\Sigma_{-}} \mathrm{d}^{d-1} \sigma \sqrt{\tilde{h}} F_{-}\left(g_{\mu \nu} ; X_{-}^\mu\right)\right]\,,
\end{aligned}
\end{equation}
where $\mathcal{M}$ is the bulk region, whose future and past boundaries are $\Sigma_{\pm}$ anchored on the  boundary time slice $\Sigma_{\mathrm{CFT}}$, with $\Sigma_{+} \cup \Sigma_- \equiv \partial \mathcal{M}$.  The codimension-one boundaries of $\mathcal{M}$ are determined  by the hypersurfaces $X_{ \pm}^\mu=0$. There are three independent scalar functions $G, F_{\pm}$. The first one can be defined by the bulk spacetime curvature invariants, and the latter two   by the curvature of $\mathcal{M}$ or the extrinsic curvatures of $\Sigma{\pm}$. Note that for simplicity  we   use the same functionals for singling out a bulk region together with its boundries and for evaluating the observables.

Considering the explicit form of the MP-AdS black hole, the functional becomes
\begin{equation}\label{fj39j39r32}
C_\mathrm{0}(\tau)=\frac{\Omega_{d-2}}{\mathrm{G}_\mathrm{N}\ell}\sum_{\varepsilon=+,-} \int_{\Sigma_{\varepsilon}}\mathrm{d}\sigma\left[h(r_{\varepsilon}) r_{\varepsilon}^{d-3} \sqrt{-f(r_{\varepsilon})^2 \dot{v}^2+2 g(r_{\varepsilon}) f(r_{\varepsilon}) \dot{v} \dot{r_{\varepsilon}}} \mathfrak{a}(r_{\varepsilon})-\varepsilon \dot{v}_{\varepsilon} b\left(r_{\varepsilon}\right)\right]
\end{equation}
where the new function $b(r)$ arises from
\begin{equation}
    \sqrt{-g}\vert_r G=\frac{\mathrm{d}b(r)}{\mathrm{d}r} 
\end{equation}
 where $ \sqrt{-g}\vert_r$ is $\sqrt{-g}$
with the angular part   removed. For the future and past codimension-one boundaries $\Sigma_\pm$, we can recognize two independent Lagrangians
\begin{equation}
\mathcal{L}_{ \pm} \equiv  h(r) r^{d-3} \sqrt{-f(r)^2 \dot{v}^2+2 g(r) f(r) \dot{v} \dot{r}} \mathfrak{a}(r)\mp\varepsilon \dot{v} b\left(r\right)\,.
\end{equation}
The codimension-one boundaries are determined by the equation of motion of a pseudoparticle, 
\begin{equation}\label{wtepid2}
f(r)^2 g(r)^2 \dot{r}^2+U_0(r)=\left(P_v^{ \pm} \pm b(r)\right)^2\,,
\end{equation}
similar to \eqref{wtepid}, where the conserved momenta $P_v^{ \pm}$ are
\begin{equation}
P_v^{ \pm}=\frac{\partial \mathcal{L}_{ \pm}}{\partial \dot{v}}=\dot{r}-f(r) \dot{v} \mp b(r)\,.
\end{equation}
The turning points $r_{\min}^{\pm}$ of the psudoparticles can be determined by
\begin{equation}
    U_0(r_{\min}^{\pm})=\left(P_v^{ \pm} \pm b(r_{\min}^{\pm})\right)^2\,.
\end{equation}
The growth rate of the generalized codimension-zero complexity can be obtained, similarly to the generalized volume complexity case, as
\begin{equation}\label{jf3p929}
\frac{\mathrm{d}C_{\mathrm{0}}(\tau)}{\mathrm{d} \tau} =\frac{\Omega_{d-2} }{\mathrm{G}_\mathrm{N}\ell}\left(P^+_v(\tau)+P^-_v(\tau)\right)\,.
\end{equation}
At late time, the growth rate becomes linear,
\begin{equation}
\lim _{\tau \rightarrow \infty}\left(\frac{\mathrm{d} C_0}{\mathrm{d} \tau}\right)=\frac{\Omega_{d-2}}{\mathrm{G}_\mathrm{N} \ell}\left( P^+_\infty+P^-_\infty\right)\,,
\end{equation}
where $P^\pm_\infty$ can be determined by similar conditions as \eqref{conmu}, explicitly shown as
\begin{equation}
\mathcal{U}_{ \pm}\left(P_{\infty}^{ \pm}, r_{f, \pm}\right)=0\,, \quad \partial_r \mathcal{U}_{ \pm}\left(P_{\infty}^{ \pm}, r_{f, \pm}\right)=0\,, \quad \partial_r^2 \mathcal{U}_{ \pm}\left(P_{\infty}^{ \pm}, r_{f, \pm}\right) \leq 0\,,
\end{equation}
where the redefined effective potential reads
\begin{equation}
\mathcal{U}_{ \pm}\left(P_v^{ \pm}, r\right)=U_0(r)-\left(P_v^{ \pm} \pm b(r)\right)^2\,.
\end{equation}

Comparing \eqref{jf3p929} with \eqref{djio289}, we can know that the growth rate of the codimension-zero complexity is determined by the sum of two momenta corresponding to two extremal surfaces, instead of only one for the codimension-one generalized volume complexity. As a result, the phase transition behaviours of the codimension-zero observable for the MP-AdS black hole can be more diverse. However, the equal area law we discussed in the previous section can still be applied   to determine the transition time of the phase transitions, if we view $P_v^{+}(\tau)+P_v^{-}(\tau)$ as a whole. Furthermore, we can infer that the result of the scaling law for the codimension-zero complexity  will not change. This can be obtained by viewing $P_v^{+}(\tau)+P_v^{-}(\tau)$ as separate terms, and each term yields a same scaling law with the generalized volume complexity. 

In \eqref{fj39j39r32}, when we set $G=0=F_-,\,F_+=1$, the CV proposal can be recovered.  The CA proposal is obtained by setting $F_{\pm}=0,\,G=R-2\Lambda$ with $\Lambda$ the cosmological constant, and if setting $F_{\pm}=0,\,G=1$, we have the CV 2.0 proposal. One thing to note is that it seems to be possible to construct a generalized CV 2.0 proposal, by choosing a generalized scalar function $G(g_{\mu\nu})$ and evaluating it on a deformed WDW patch. However, we will not pursue the details of this potential generalized CV 2.0 proposal, leaving it for future study.

\section{Conclusions and discussion}\label{sec:clr}

In this paper, we studied the generalized holographic complexity of rotating black holes. Our work was inspired by recent studies of the construction of generalized codimension-one volume complexity and codimension-zero complexity \cite{Belin:2021bga, Belin:2022xmt}, the discovery of complexity phase transitions \cite{Wang:2023eep,Jorstad:2023kmq}, and the scaling relations of the complexity of formation for higher-dimensional rotating black holes \cite{AlBalushi:2020rqe, AlBalushi:2020heq}. We aimed to study the generalized holographic complexity of rotating boundary states which have a dual bulk geometric description in terms of higher-dimensional rotating MP-AdS black holes. We put our emphasis on studying the generalized codimension-one volume complexity and then extended the investigation to the codimension-zero observable. Our conclusions can be summarized as follows.

First, we showed that at late times the generalized holographic complexity grows linearly, complying with the requisite property of holographic complexity dual to rotating TFDs on the two-sided boundary CFTs. Specifically, from the generalized volume functional \eqref{volfunc}, we found the extremal surface determined by \eqref{dotr8} and \eqref{dotr82}. According to the definition \eqref{codoneo} of generalized volume complexity, we demonstrated that its growth rate becomes constant at late times, which corresponds to a specific codimension-one hypersurface defined by the conserved momentum \eqref{constc2} satisfying the condition \eqref{conmu}, where the boundary time tends to infinity. The result thus implies that the generalized holographic complexity defined through a functional constructed from diffeomorphism invariants is well-posed for higher-dimensional stationary black holes (at least for the odd-dimensional MP-AdS black holes).

Second, we studied phase transition behaviour of the generalized CV complexity at early times. We found that the inclusion of angular momentum introduces qualitatively new features in generalized volume complexity, resulting in several new phase transitions. We specified the scalar function \( F \) in \eqref{volfunc} to be \eqref{fonelg}, which is defined by the covariant Gauss-Bonnet invariant. Then, we studied variation of the complexity over time. We found that there can be several types of new transitions in the growth of complexity at early times, dependent on the relative magnitudes of the maxima for the effective potential. Specifically, with the angular momentum and other parameters finely tuned, the complexity can begin in an initial phase 1 and then either transit to phase 3 (type (A)) or to phase 2 and then to phase 3 (type (B)) if the maxima are relatively comparable in size, with \( U_{0}(r_L) \gtrsim U_{0}(r_M) \gtrsim U_{0}(r_R) \). If \( U_{0}(r_L) \gtrsim U_{0}(r_M) \gg U_{0}(r_R) \), we found that phase 1 is inadmissible, and the generalized volume complexity begins at phase 2 and ends at phase 3 (type (C)). Finally, if \( U_{0}(r_L) \gg U_{0}(r_M) \gtrsim U_{0}(r_R) \), the complexity experiences only the single growth phase 3 (type (D)). Regardless of which type takes place, all generalized volume complexities grow linearly over time at late times in the last phase.
 
Our study of the phase transitions for the generalized holographic complexity extends  previous results for the three-dimensional rotating BTZ black hole to the higher-dimensional rotating case of the odd-dimensional MP-AdS black hole with equal angular momenta, and serves as a comparison with its Schwarzschild-AdS and RN-AdS counterparts. The novel behaviours of the generalized volume complexity, which primarily originate from the curvature invariants of the MP-AdS black holes, are independent of the number of spacetime dimensions and cannot be found in the higher-dimensional Schwarzschild-AdS or RN-AdS black holes. However higher-order Lovelock terms or higher-order Weyl square terms for the Schwarzschild-AdS black holes admit similar phase behaviour, as recently demonstrated \cite{Jorstad:2023kmq, Jiang:2023jti}, though not as complicated as the case we have studied here. Moreover, our investigation further confirms that generalized holographic complexity is a quantity quite similar to the thermodynamic free energy. A thermodynamic system tends to equilibrate  at a lower free energy state. By contrast,  from the definition of the generalized holographic complexity, the TFDs on the boundary CFTs also tend to be in a state with higher holographic complexity, and thus abundant global pseudo-phase behaviour can occur.

We also found that phase transitions of   generalized holographic complexity satisfy an area law.  This law is based on the fact that the phase transition  of the complexity is first-order. This Maxwell-like area law has different properties  compared to Maxwell's equal area law for thermodynamic phase transitions of black holes \cite{Kubiznak:2016qmn}. This is a generalized area insofar as  areas that are related with infinite boundary time are modified by subtracting a constant determined by the difference of the generalized holographic complexity at left and right infinite boundary time around the divergent point.  Moreover, the area law applies to a chronological variation process instead of an equilibrium thermodynamic process. Note that this area law seems insofar the only one discovered for the non-thermodynamic process. The area law may involve an equality relation of more than two regions in the diagram for the variation of boundary time with respect to the conserved momentum of the extremal surface. These multiple regions appear due to the nontrivial contribution from the rotation of the black hole. Based on this  area law, we showed that there exists a second-order phase transition at a critical point for the chronological variation of the complexity, where the areas  enclosed by the boundary time curves as well as the complexity differences at the infinite boundary time vanish. 

More interestingly, by the area law we also illustrated that there can be a multi-critical point for the complexity phase transition, where two first-order phase transition points intersect at a single point. This is reminiscent of the multi-critical thermodynamic phase transitions for the black holes first discovered in \cite{Altamirano:2013uqa,Wei:2014hba,Tavakoli:2022kmo}. We obtained a phase diagram qualitatively exhibiting the phase structures for generalized complexity. The phase diagram manifests the nontrivial contribution of the MP-AdS black hole's  angular momentum to the phase transitions of  generalized holographic complexity. 

Evidently complexity phase transitions   are determined by the choice of the scalar function that defines the generalized holographic complexity itself. The inclusion of   angular momentum   makes the   transitions somewhat more complicated. In the generalized holographic complexity proposal, the boundary TFD states correspond to the generalized holographic complexity of the bulk rotating black hole spacetime. Under this conjecture, the complicated phase transition behaviours of the complexity also indicate the phase transitions of the quantum complexity of the rotating TFD states. However, it is still an open question as to how to identify these phase transitions in the boundary quantum theories.

In considering  the effects of  angular momentum on  complexity further,  we defined the generalized holographic complexity of formation and explored its scaling behaviour. We discovered that the generalized holographic complexity of formation for the large near-extremal MP-AdS black hole   still scales with the  thermodynamic volume of the black hole, 
 provided  \eqref{fji39pzl} is satisfied so that a generalized holographic complexity of formation can be  defined.  We further verified  the scaling relation of the generalized complexity of formation for the large near-extremal RN-AdS black hole, and found that the  scaling behaviour is not determined by the black hole entropy if the scalar function being divergent at spatial infinity (cf. \eqref{jf39q8qnz}).

These studies on the generalized holographic complexity of formation may serve as judgment on the scaling behaviours of the generalized holographic complexity. For the MP-AdS black hole case, we pointed out that if the condition \eqref{fji39pzl} is not satisfied by the scalar function $F$ that defines the generalized holographic complexity in \eqref{codoneo}, the generalized holographic complexity of formation will not be well-defined.  For the RN-AdS black hole, its scaling relation will be commensurate with the original one only if \eqref{jf39q8qnz} is not  satisfied. As a result, we know that if the  definition  of the generalized holographic complexity \eqref{codoneo} satisfies the conditions
\begin{align}
     F\left(g_{\mu \nu}, \mathcal{R}_{\mu \nu \rho \sigma}, \nabla_\mu\right)&<\infty\,,  \quad\mathrm{in\,\, vacuum\,\, AdS\,\,geometry\,,}\\
F\left(g_{\mu \nu}, \mathcal{R}_{\mu \nu \rho \sigma}, \nabla_\mu\right)&<\infty\,,\quad \mathrm{at\,\, spatial\,\, infinity\,,}
\end{align}
the scaling law of the generalized complexity of formation will be the same as that of the original complexity of formation; otherwise, either the generalized complexity of formation does not exist or the scaling laws will be dependent on the distinct generalized holographic complexities.

Our results imply the nontrivial holographic properties of the higher-dimensional rotating MP-AdS black holes and also exhibit the universal property of the generalized holographic complexity. We have briefly shown that the above discoveries also apply to the codimension-zero generalized complexity. One challenge for future study may be to consider the generalized CV 2.0 proposal in the sense of calculating the spacetime volume of the deformed WDW patch. In addition, verifying the switchback effect of the generalized holographic complexity for the rotating boundary states remains a task for future exploration.

\acknowledgments

This work was supported by the Natural Sciences and Engineering Research Council of Canada and the National Natural Science Foundation of China (Grants No. 12365010, No. 12005080, and No. 12064018). M. Z. was also supported by the Chinese Scholarship Council Scholarship (No. 202208360067).
\appendix
\section{Some curvature invariants}\label{app:cio}

We will present the curvature invariants for the five-dimensional and seven-dimensional MP-AdS black holes with equal angular momenta. In the five-dimensional case, we have the following curvature invariants
\begin{equation}
\mathcal{R}^2=\frac{400}{\ell^4}\,,\quad\mathcal{R}_{\mu\nu } \mathcal{R}^{\mu\nu }=\frac{80}{\ell ^4}\,,
\end{equation}
\begin{equation}
\begin{aligned}
&\mathcal{R}_{\mu\nu\rho\sigma} \mathcal{R}^{\mu\nu\rho\sigma}\\&=\frac{8 \left(12 a^4 m^2 \left(3 r^4+16 r^2 \ell ^2+16 \ell ^4\right)-24 a^2 m^2 r^2 \ell ^2 \left(3 r^2+8 \ell ^2\right)+r^4 \left(36 m^2 \ell ^4+5 r^8\right)\right)}{r^{12} \ell ^4}\,,
\end{aligned}
\end{equation}
\begin{equation}
\mathcal{C}^{abcd}\mathcal{C}_{abcd}=\frac{96 m^2 \left(a^4 \left(3 r^4+16 r^2 \ell ^2+16 \ell ^4\right)-2 a^2 \left(3 r^4 \ell ^2+8 r^2 \ell ^4\right)+3 r^4 \ell ^4\right)}{r^{12} \ell ^4}\,,
\end{equation}
\begin{equation}
\mathcal{G}=\frac{24 \left(4 a^4 m^2 \left(3 r^4+16 r^2 \ell ^2+16 \ell ^4\right)-8 a^2 m^2 r^2 \ell ^2 \left(3 r^2+8 \ell ^2\right)+r^4 \left(12 m^2 \ell ^4+5 r^8\right)\right)}{r^{12} \ell ^4}\,,
\end{equation}
where $\mathcal{C}_{abcd}$ is the Weyl tensor. Similarly, for the seven-dimensional case, we have the following curvature invariants
\begin{equation}
\mathcal{R}^2=\frac{1764}{\ell^2}\,,\quad\mathcal{R}_{\mu\nu } \mathcal{R}^{\mu\nu }=\frac{252}{\ell ^4}\,,
\end{equation}
\begin{equation}
\begin{aligned}
\mathcal{R}_{\mu\nu\rho\sigma} \mathcal{R}^{\mu\nu\rho\sigma}&=\frac{12 \left(-16 a^2 m^2 r^2 \ell ^2 \left(25 r^2+48 \ell ^2\right)+200 m^2 r^4 \ell ^4+7 r^{16}\right)}{r^{16} \ell ^4}\\&\quad+\frac{12 \left(8 a^4 m^2 \left(25 r^4+96 r^2 \ell ^2+80 \ell ^4\right)\right)}{r^{16} \ell ^4}\,,
\end{aligned}
\end{equation}
\begin{equation}
\mathcal{C}^{\mu\nu\rho\sigma}\mathcal{C}_{\mu\nu\rho\sigma}=\frac{96 m^2 \left(a^4 \left(25 r^4+96 r^2 \ell ^2+80 \ell ^4\right)-2 a^2 \left(25 r^4 \ell ^2+48 r^2 \ell ^4\right)+25 r^4 \ell ^4\right)}{r^{16} \ell ^4}\,,
\end{equation}
\begin{equation}
\mathcal{G}=\frac{24 \left(4 a^4 m^2 \left(25 r^4+96 r^2 \ell ^2+80 \ell ^4\right)-8 a^2 m^2 r^2 \ell ^2 \left(25 r^2+48 \ell ^2\right)+5 r^4 \left(20 m^2 \ell ^4+7 r^{12}\right)\right)}{r^{16} \ell ^4}\,.
\end{equation}
It is clear that all these curvature invariants of the MP-AdS black holes are independent of the angular coordinates. Additionally, it can be confirmed that the difference between the squared Weyl tensor and the Gauss-Bonnet invariant remains constant, whether for five-dimensional or seven-dimensional MP-AdS black holes.

\section{Growth rate of complexity}\label{dcdtaud}

The generalized complexity $C$ can be expressed as
\begin{equation}
\begin{aligned}
C & =\frac{1}{\mathrm{G}_\mathrm{N} \ell} \int_{\Sigma} \mathrm{d}^{d-1} \sigma \sqrt{\tilde{h}} F\left(g_{\mu \nu}, \mathcal{R}_{\mu \nu \rho \sigma}, \nabla_\mu\right) \\
& =\frac{\Omega_{d-2}}{\mathrm{G}_\mathrm{N} \ell} \int_{\Sigma} \mathrm{d} \sigma \mathcal{L}(r,\dot{r},\dot{v})\,,
\end{aligned}
\end{equation}
where $\mathcal{L}(r,\dot{r},\dot{v})$ is the Lagrangian. We have the derivative of the Lagrangian $\mathcal{L}$ with respect to $\dot{t}$, which is given by
\begin{equation}\label{app1}
\frac{\partial \mathcal{L}}{\partial \dot{t}}=\frac{\partial \mathcal{L}}{\partial \dot{v}} \frac{\partial \dot{v}}{\partial \dot{t}}=\frac{\partial \mathcal{L}}{\partial \dot{v}}\,.
\end{equation}
The rate of change of the complexity with respect to the boundary time $\tau$ can be calculated as
\begin{equation}
\begin{aligned}
\frac{\mathrm{d} C}{\mathrm{d} \tau} & =\frac{\Omega_{d-2}}{\mathrm{G}_\mathrm{N} \ell} \int_{\Sigma} \mathrm{d} \sigma \frac{\mathrm{d} \mathcal{L}}{\mathrm{d} \tau}  =\frac{\Omega_{d-2}}{2 \mathrm{G}_\mathrm{N} \ell} \int_{\Sigma} \mathrm{d} \sigma \frac{\mathrm{d} \mathcal{L}}{\mathrm{d} t} =\frac{\Omega_{d-2}}{2 \mathrm{G}_\mathrm{N} \ell} \int_{\Sigma} \mathrm{d} \sigma \frac{\mathrm{d}}{\mathrm{d} \sigma} \frac{\partial \mathcal{L}}{\partial \dot{t}} \\&=\frac{\Omega_{d-2}}{2\mathrm{G}_\mathrm{N} \ell}\int_{\Sigma} \mathrm{d}\left(\frac{\partial \mathcal{L}}{\partial \dot{t}}\right)=\frac{\Omega_{d-2}}{2\mathrm{G}_\mathrm{N} \ell}\int_{\Sigma} \mathrm{d}\left(\frac{\partial \mathcal{L}}{\partial \dot{v}}\right)=\frac{\Omega_{d-2}}{2\mathrm{G}_\mathrm{N} \ell}\left.\mathcal{P}_v\right|_{\partial \Sigma(\tau)}=\frac{\Omega_{d-2}}{\mathrm{G}_\mathrm{N} \ell}P_v (\tau)\,,
\end{aligned}
\end{equation}
where $\mathcal{P}_v$ is the conjugate momentum of $v$ and in the second line, we have used \eqref{app1}.

\begin{figure}[t!]
	\centering
	\includegraphics[width=3.5in]{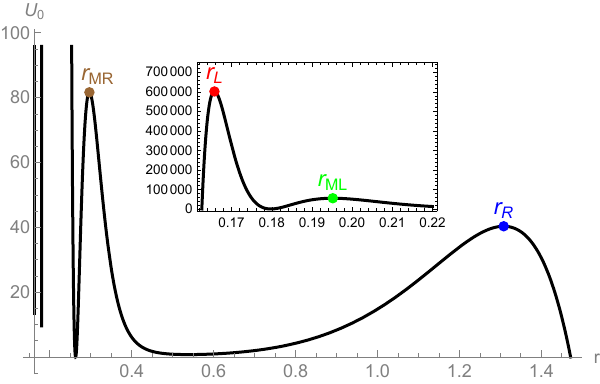}
	\caption{The effective potential $U_0$ between the inner and outer horizons for the seven-dimensional MP-AdS black hole with $m=9, a=4/25, \ell=9/10, \lambda=1/10^{171/20}$ (these parameter values applies to Figs. \ref{taup7}, and \ref{ctau7}). The effective potential has local extrema at $r=r_L, r_{ML}, r_{MR}, r_R$.}\label{effpmp7}
\end{figure}

\begin{figure}[t!]
	\centering
	\includegraphics[width=5in]{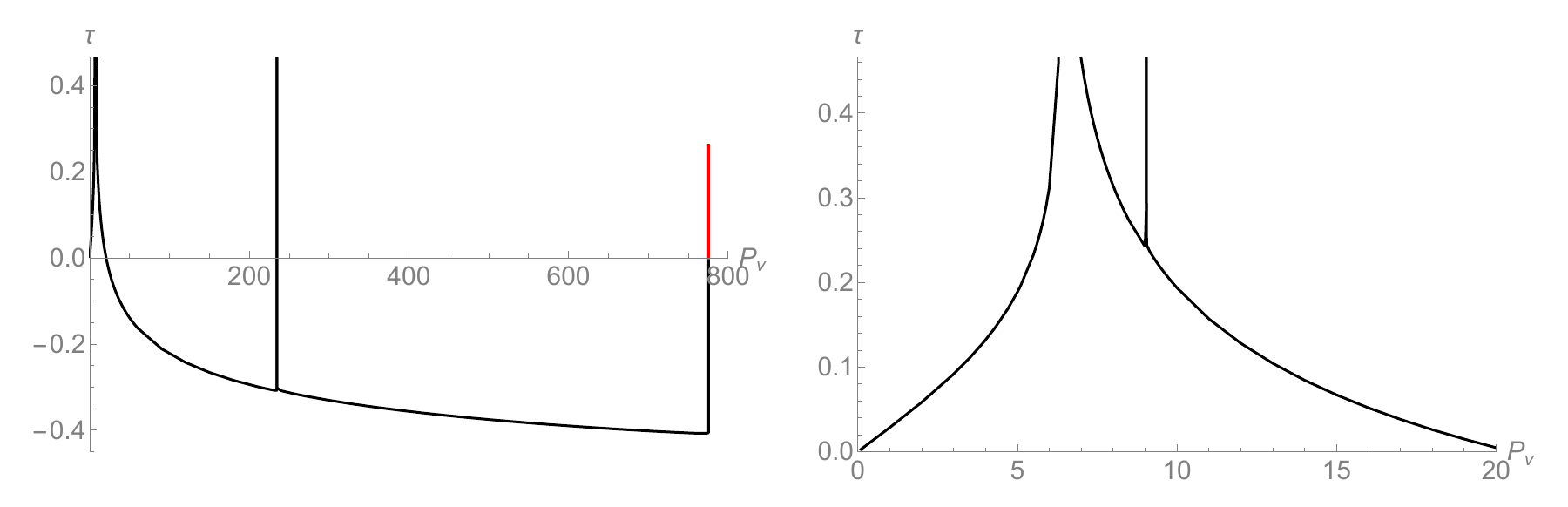}
	\caption{Left: the variation of the boundary time with respect to the conserved momentum for the seven-dimensional MP-AdS black hole, with the real path for the time variation denoted by the red curve.  Right: details of $\tau$-variation in the small $P_v$ range.}\label{taup7}
\end{figure}

\begin{figure}[t!]
	\centering
	\includegraphics[width=5in]{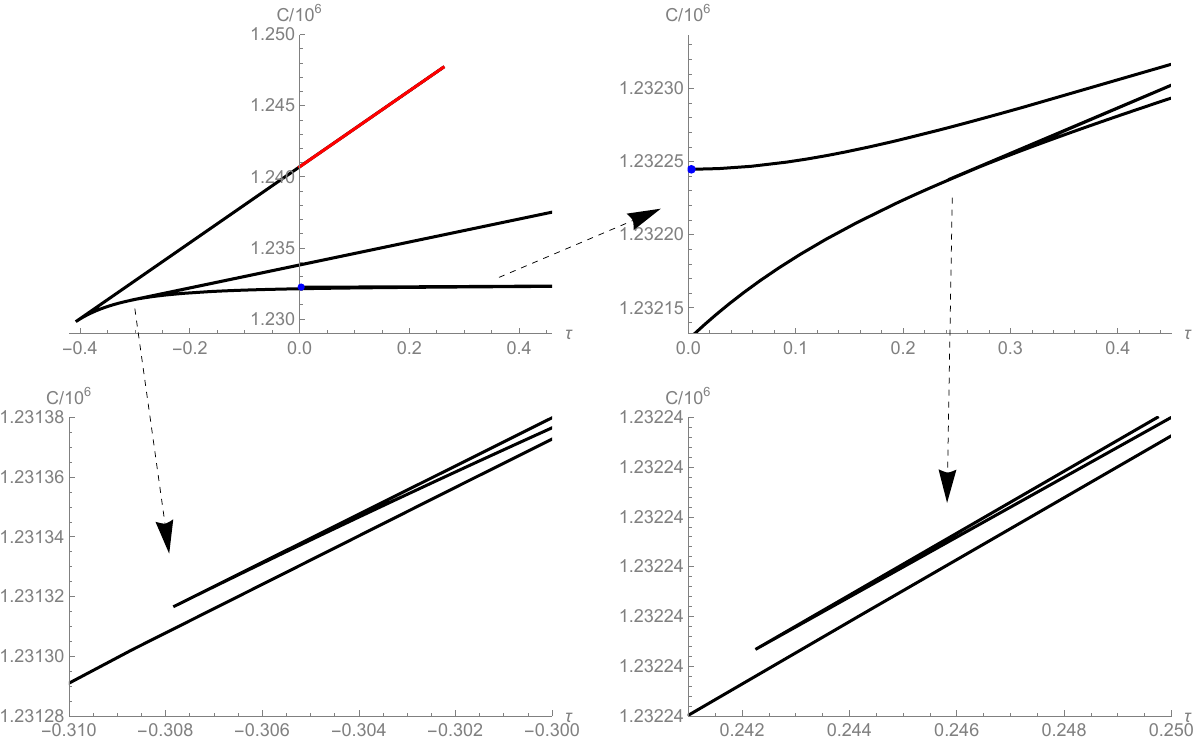}
	\caption{The variation of the generalized volume complexity with respect to the boundary time. We can read the variation of the complexity from the blue point. There is only one phase 4, whcih is the real variation path of the complexity, as represented by the red curve.}\label{ctau7}
\end{figure}

\section{Seven-dimensional case}\label{app:seven}

We show here one typical variation of the generalized volume complexity for the seven-dimensional MP-AdS black hole. We study the case where there are four extremal maxima in the effective 
potential\footnote{We note that there can also be up to four extremal maxima for the effective potential in the five-dimensional MP-AdS black hole background.}, as shown in Fig. \ref{effpmp7}. The extremal maxima  $r_L, r_{ML}, r_{MR}, r_R$
of the effective potential respectively  yield four critical momenta $P_L, P_{ML}, P_{MR}, P_R$;  the boundary time diverges at each, as shown in Fig. \ref{taup7}. After calculating the variation of the complexity with respect to the boundary time, as graphically depicted in Fig. \ref{ctau7}, we see that there can be only one phase for the real path of the complexity variation, denoted by the red curve. This is consistent with the type (D)-variation of the complexity shown in Fig. \ref{effpmp4}. The direct reason for this type of complexity variation path is understandable, as according to the parameters we have chosen for the effective potential in Fig. \ref{effpmp7}, we have $U_0 (r_L)$ being far greater than the other three extremal maxima $U_0 (r_{ML}), U_0 (r_{MR}), U_0 (r_{R})$. Of course, by tuning the parameters, we can obtain proper configurations for the effective potential that produce more complicated types of complexity variations, as we have done in five dimensions for the three extremal maxima cases in Figs. \ref{effpmp1}, \ref{effpmp2}, \ref{effpmp3}, and \ref{effpmp4}. 

 Obtaining specific values of $m,\; a,\; \ell,\; \lambda$ is computationally quite tedious due to the large powers of the radial variable and the presence of only  three free parameters (we can rescale $m,\; a$ by $\ell$).  However
the requirement of an appropriate number and size of maxima and minima of $U_0$ yields  more than three restrictions on these variables. In general it is not clear that any desired form of $U_0$ can be obtained for the specific choice 
\eqref{fonelg} 
of generalized complexity we are considering.  We have found a numerical scan of the parameters to be most effective in exploring the different possibilities, but have not found any nontrivial behaviour beyond what we have presented in this paper.

\bibliographystyle{jhep}
\bibliography{refs}

\end{document}